\documentclass[twocolumn,amsmath,amssymb,floatfix,footinbib,superscriptaddress]{revtex4-2}
\usepackage{natbib,bm,graphicx,url,epsf,xcolor}
\usepackage[british]{babel}
\usepackage{wasysym}
\usepackage{MnSymbol}
\setcitestyle{super} 
\usepackage{soul} 
\usepackage{enumitem}
\usepackage{braket,setspace}

\usepackage{titlesec}
\titlespacing{\section} {-10pt}{3ex}{2ex}
\titleformat{\section}
{\centering\normalfont\small\bfseries\MakeUppercase}{Section~\thesection}{-10ex}{}

\graphicspath{{Figures/}} 

\usepackage{pdfpages} 
\usepackage{pgffor} 

\makeatletter
\AtBeginDocument{\let\LS@rot\@undefined}
\makeatother

\newif\ifarXiv
\arXivtrue

\begin{document}
\title{Experimental Evidence of Fractional Entropy in Critical Kondo Systems}

\author{C. Piquard}
\email[e-mail: ]{colin.piquard@espci.org}
\affiliation{Universit\'e Paris-Saclay, CNRS, Centre de Nanosciences et de Nanotechnologies, 91120 Palaiseau, France}
\author{A. Veillon}
\affiliation{Universit\'e Paris-Saclay, CNRS, Centre de Nanosciences et de Nanotechnologies, 91120 Palaiseau, France}
\author{Y. Sato}
\affiliation{Universit\'e Paris-Saclay, CNRS, Centre de Nanosciences et de Nanotechnologies, 91120 Palaiseau, France}
\author{F. Zanichelli}
\affiliation{Universit\'e Paris-Saclay, CNRS, Centre de Nanosciences et de Nanotechnologies, 91120 Palaiseau, France}
\author{A. Aassime}
\affiliation{Universit\'e Paris-Saclay, CNRS, Centre de Nanosciences et de Nanotechnologies, 91120 Palaiseau, France}
\author{A. Cavanna}
\affiliation{Universit\'e Paris-Saclay, CNRS, Centre de Nanosciences et de Nanotechnologies, 91120 Palaiseau, France}
\author{U. Gennser}
\affiliation{Universit\'e Paris-Saclay, CNRS, Centre de Nanosciences et de Nanotechnologies, 91120 Palaiseau, France}
\author{A. K. Mitchell}
\affiliation{School of Physics, University College Dublin, Belfield, Dublin 4, Ireland}
\affiliation{Centre for Quantum Engineering, Science, and Technology, University College Dublin, Ireland}
\author{A. Anthore}
\email[e-mail: ]{anne.anthore@c2n.upsaclay.fr}
\affiliation{Universit\'e Paris-Saclay, CNRS, Centre de Nanosciences et de Nanotechnologies, 91120 Palaiseau, France}
\affiliation{Universit\'{e} Paris Cit\'{e}, CNRS, Centre de Nanosciences et de Nanotechnologies, F-91120 Palaiseau, France}
\author{F. Pierre}
\email[e-mail: ]{frederic.pierre@cnrs.fr}
\affiliation{Universit\'e Paris-Saclay, CNRS, Centre de Nanosciences et de Nanotechnologies, 91120 Palaiseau, France}


\begin{abstract}
\noindent Unconventional quantum states defying the ubiquitous Fermi-liquid paradigm can emerge in the presence of strong electronic correlations. 
Among these, non-Abelian anyons—such as Majorana zero modes and Fibonacci anyons—are of particular interest for topological quantum computing due to their non-integer quantum dimensions $d>1$, which allows for protected non-local encoding and processing of quantum information\cite{nayak_non-abelian_2008}.
However, despite considerable efforts, the unambiguous characterisation of such anyons via transport measurements has proved challenging\cite{Yazdani_huntMZM_2023}.
Instead, here we provide experimental evidence for the low-temperature fractional entropy $\Delta S$ associated with a single anyon, which directly implies its non-Abelian character through the relation $\Delta S=k_\mathrm{B}\ln d$.
This thermodynamic signature\cite{nayak_non-abelian_2008,yang_thermopower_2009,SelaMZM2019,Kleeorin_S-ThermoElec_2019} is measured in metal-semiconductor quantum circuits engineered to realize quantum-critical states from frustrated interactions.
Using a micrometre-scale metallic island coupled to two or three electronic leads, we tune the system to two-channel and three-channel Kondo critical points\cite{Matveev1991,Matveev1995,Iftikhar2015,Iftikhar2018}. 
By measuring the island charge and exploiting a thermodynamic Maxwell relation\cite{Hartman2018}, we estimate the entropy associated with the anyons that emerge in these critical states.  
Our observations reveal fractional values, exposing non-Abelian anyons. The corresponding scaling dimensions are consistent with theoretical predictions for a Majorana zero mode $\Delta S=k_\mathrm{B}\ln\sqrt{2}$ and a Fibonacci anyon $\Delta S=k_\mathrm{B}\ln\tfrac{1}{2}(1+\sqrt{5})$ for two and three channels\cite{affleck1991universal,Emery1992,lopes2020anyons}.
These findings establish entropy measurements as a powerful tool for characterizing exotic quantum states.
\end{abstract}
\maketitle

Entropy is a cornerstone of thermodynamics and a powerful lens through which to explore quantum matter. 
It can reveal hidden features such as fractionalized excitations\cite{nayak_non-abelian_2008,yang_thermopower_2009,cooper_observable_2009,SelaMZM2019}, emergent degrees of freedom\cite{saito_isospin_2021,rozen_entropic_2021} and topological order\cite{kitaev_topological_2006,fendley_topological_2007,sankar_measuring_2023}. 
The ground state entropy is a particularly discerning quantity for systems featuring non-Abelian anyons\cite{yang_thermopower_2009,cooper_observable_2009,SelaMZM2019,Kleeorin_S-ThermoElec_2019}. 
Braiding multiple non-Abelian anyons leads to a transformation between degenerate many-body ground states, a process that underpins proposals for topological quantum computation\cite{nayak_non-abelian_2008} due to its non-local nature, protecting logical qubit states from local perturbations.
This process also directly translates into a high ground state degeneracy $d^N$ for systems with $N$ non-Abelian anyons, where $d$ is their so-called quantum dimension\cite{nayak_non-abelian_2008,fendley_topological_2007}.
In fact, $d>1$ is the defining signature of non-Abelian statistics\cite{rowell_degeneracy_2016}, whereas fermions and other Abelian anyons have $d=1$. In sharp contrast with a standard degeneracy, the quantum dimension can be non-integer.
For example $d=\sqrt{2}$ for Majorana zero modes and $d=\phi$ for Fibonacci anyons, with $\phi=\tfrac{1}{2}(1+\sqrt{5})$ the golden ratio.
Thus, when adding one non-Abelian anyon, a distinct thermodynamic signature is obtained through the ground state entropy increase by the fractional value $k_\mathrm{B}\ln d$\cite{nayak_non-abelian_2008,yang_thermopower_2009,SelaMZM2019,Kleeorin_S-ThermoElec_2019}.

A practical challenge is that the contribution of a single quasiparticle to the ground state entropy of a bulk system is often relatively small in finite-temperature experiments. 
Instead one can study the so-called ``impurity'' entropy $S_\mathrm{imp}$, corresponding to the \textit{change} in total entropy associated with introducing or tuning an impurity\cite{Wilson1975,Vojta2006}. 
Importantly, the impurity entropy can be obtained experimentally from thermodynamic Maxwell relations with outstanding precision\cite{child_entropy_2022,adam_entropy_2025}, better than the relevant scale $k_\mathrm{B}\ln2$. With this approach, the quantum states to be investigated must be induced by the presence of the impurity -- as with the nontrivial boundary critical states arising at impurity quantum phase transitions\cite{Vojta2006}.

A paradigmatic example is the multi-channel Kondo effect\cite{Cox1998, Nozieres1980, affleck1991universal,Vojta2006}, which develops when multiple independent electronic baths (or ``channels'') compete to screen a single spin-$\tfrac{1}{2}$ impurity. As illustrated in Fig.~\ref{fig1}a for the two-channel setup, a quantum-critical point arises when the channels couple equally, thereby frustrating the standard Kondo screening process. This results in non-Fermi-liquid behavior in the vicinity of the quantum phase transition, with the impurity being only partially screened. Its effective degeneracy $d$ is reduced from 2 (for a free spin-$\tfrac{1}{2}$ degree of freedom) to a non-integer value\cite{affleck1991universal} $1<d<2$, which is the quantum dimension of an unscreened non-Abelian anyon\cite{lopes2020anyons} remaining at the impurity site. The residual impurity entropy at zero temperature therefore takes a characteristic value $S_\mathrm{imp}=k_\mathrm{B}\ln d$.

Specifically, a Majorana zero mode, bound to the impurity site, is predicted at low temperatures in the two-channel Kondo (2CK) setup\cite{Emery1992} giving $S_\mathrm{imp}=k_\mathrm{B}\ln \sqrt{2}$, whereas at the three-channel Kondo (3CK) critical point\cite{lopes2020anyons}, $S_\mathrm{imp}=k_\mathrm{B}\ln \phi$ corresponds to the additional entropy of a Fibonacci anyon.
Indeed, frustrated Kondo interactions generally produce criticality and lead to fractionalized anyonic quasiparticles\cite{affleck1991universal,Emery1992,Vojta2006,lopes2020anyons,karki2023z}. 
Multi-channel Kondo systems have therefore become a prominent playground for exploring non-Fermi-liquid physics and criticality, driven by experimental demonstrations that such phenomena can be controllably engineered in tunable quantum circuits\cite{Potok2007,Keller2015,Mebrahtu2012,Mebrahtu2013,Iftikhar2015,Iftikhar2018,Pouse_2IKcharge_2022}.

Here, we report on the experimental observation of a fractional impurity entropy in the 2CK and 3CK states.  
These are implemented using a charge-Kondo architecture\cite{Matveev1991,Matveev1995,LeHur2004,Nguyen2018seebeck2ICK} based on a metallic island within a hybrid metal-semiconductor circuit\cite{Iftikhar2015,Iftikhar2018,Pouse_2IKcharge_2022,piquard_observing_2023}. 
Previous studies relying on transport measurements\cite{han_fractional_2022} provided a self-consistent estimation of the entropy, but were necessarily \textit{indirect} and presupposed the critical Kondo state being probed. By contrast, here we make a direct observation by incorporating a charge detection system into our experimental setup, which allows us to extract the entropy change associated with the development or destruction of the critical state, via a thermodynamic Maxwell relation. 
Importantly, this approach does not presuppose anything about an underlying model, and so a measured entropy $S_{\rm imp}=k_{\rm B}\ln d$ with non-integer $d$ is an unbiased indication of the existence of a non-Abelian anyon in our setup.
By studying the universal entropic crossovers both towards and away from the critical points, the impurity entropy associated with the predicted Kondo anyons~\cite{affleck1991universal,lopes2020anyons} is bounded from above and below, hence establishing its fractional nature. Our findings are consistent, at experimental accuracy, with a free Majorana in the 2CK setup and a Fibonacci anyon for 3CK.

\section*{Multi-channel charge Kondo circuits}

\begin{figure}[ht!]
        \centering
        \includegraphics[width=85mm]{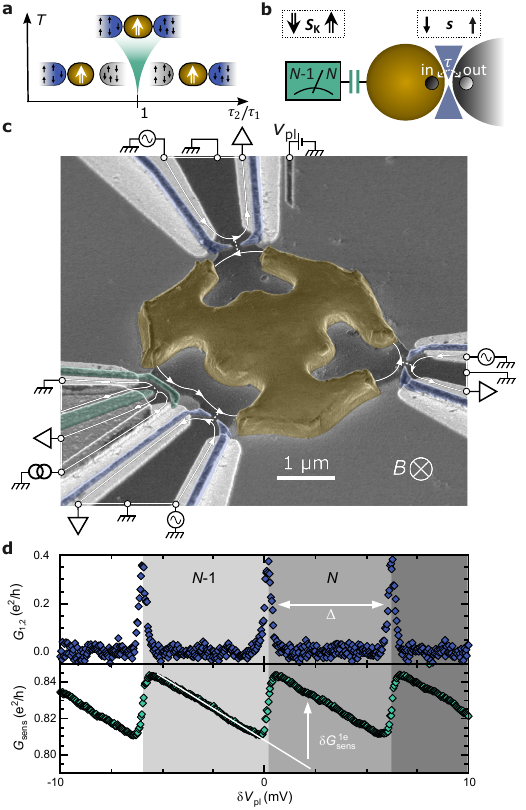}
        \caption{
        \footnotesize
        \textbf{Charge-Kondo circuit.}
        (a) Illustration of the quantum phase transition arising in two-channel Kondo systems as a function of coupling asymmetry.
        As the temperature $T$ increases, quantum criticality (green) starts at symmetric couplings ($\tau_1=\tau_2$), extends in $\tau_2/\tau_1$, and then vanishes. 
        Baths colored grey progressively decouple from the Kondo impurity as $T\rightarrow0$.
        (b) Schematic of the charge-Kondo mapping and charge sensing.
        The impurity pseudospin $\boldsymbol{S}_\mathrm{K}$ (double arrows) encoding two charge states $N-1$ and $N$ of a metallic island (gold disk) flips as each electron tunnels across a QPC of transmission $\tau$, together with the pseudospin $\boldsymbol{s}$ encoding the electron position (single arrows). Illustrated here for one Kondo channel, assuming polarized real spin.
        (c) Colored e-beam micrograph of the measured device. 
        A micron-scale metallic island (gold) is connected through tunable QPCs formed using split gates (blue) in a buried 2DEG (darker gray). 
        The island’s charge is controlled with the plunger gate voltage $V_\mathrm{pl}$ (top), and measured with a capacitively coupled sensor separated by a `barring' gate (green).
        The current propagates along the edges of the 2DEG set in the integer quantum Hall regime ($B\simeq5.3$\,T, filling factor $\nu=2$), as depicted by white lines with arrows (inner channel not shown).
        (d) The sensor calibration with a weakly connected island is obtained from the amplitude $\delta G_\mathrm{sens}^\mathrm{1e}$ of the periodic jumps in the charge sensor conductance (bottom panel) resulting from the addition of one electron at a time when sweeping $V_\mathrm{pl}$, as attested by the simultaneous conductance peaks across the Kondo device (top panel).  
        }
        \label{fig1}
\end{figure}

\noindent The multi-channel Kondo model\cite{Nozieres1980} involves a local spin-$\tfrac{1}{2}$ ``impurity'' $\mathbf{S}_\mathrm{K}$ coupled antiferromagnetically to $M>1$ baths of itinerant electrons. Its Hamiltonian reads\cite{Cox1998},
\begin{eqnarray}
H_\mathrm{MCK}=\sum_{i=1}^M J_i\left (\hat{S}_\mathrm{K}^+ \hat{s}_i^- +\hat{S}_\mathrm{K}^-\hat{s}_i^+ +\delta\,\hat{S}_\mathrm{K}^z\hat{s}_i^z\right ) +H_\mathrm{baths}
\label{eq:MCKmodel}
\end{eqnarray}
where $\hat{S}_\mathrm{K}^{\pm}$ and $\hat{s}_i^{\pm}$ are, respectively, the raising/lowering operators for the impurity spin and for the spin density of the itinerant electrons at the impurity position of bath  $i$, whereas $\hat{S}_\mathrm{K}^{z}$ and $\hat{s}_i^{z}$ are their corresponding spin projections. $H_\mathrm{baths}$ describes the $M$ independent baths. The exchange coupling $J_i>0$ favours the formation of a spin-singlet state between the impurity and bath $i$. The magnetic frustration resulting when $J_i\equiv J$ is the same for all baths is the fundamental origin of the quantum-critical point in this model. The spin anisotropy $\delta$ is irrelevant to the low-temperature physics\cite{Nozieres1980,Cox1998}. 

Pioneer implementations of a Kondo impurity in circuits\cite{Cronenwett1998,Goldhaber-Gordon1998a,Nygard2000} involved a small quantum dot with a spin-degenerate level, populated by one electron (either up or down spin). The effective model for such a setup is Eq.~\eqref{eq:MCKmodel} with $M=1$ and $\delta=1$. However the generalization to multiple Kondo channels proved challenging\cite{Potok2007,Keller2015}. 

In the charge-Kondo circuits considered in this work, the impurity is instead formed by two degenerate macroscopic quantum charge states of a metallic island\cite{Matveev1991, Matveev1995,Iftikhar2015,LeHur2004,Nguyen2018seebeck2ICK,Iftikhar2018,Pouse_2IKcharge_2022,piquard_observing_2023}, see Fig.~\ref{fig1}b. With polarized real spin, we map island charge states with $N-1$ and $N$ electrons to the pseudospin states $|\Downarrow\rangle$ and $|\Uparrow\rangle$. An electron tunneling onto the island at one of the quantum point contacts (QPCs) flips the pseudospin, $\hat{S}_\mathrm{K}^+\equiv|\Uparrow\rangle\langle\Downarrow|=|N\rangle\langle N-1|$, and similarly for $\hat{S}_\mathrm{K}^-$ which corresponds to tunneling off the island. The localization of the electron either side of QPC $i$ maps to pseudospin $s_i^z=\pm\tfrac{1}{2}$, such that the tunneling process across the QPC flips the pseudospin ($\hat{s}_i^-$ when tunneling onto the island, $\hat{s}_i^+$ when tunneling off). Overall, each QPC $i$ contributes a term like $J_i(\hat{S}_\mathrm{K}^+\hat{s}_i^-+\hat{S}_\mathrm{K}^-\hat{s}_i^+)$ where $J_i$ is controlled by the QPC transmission probability $\tau_i$. Thus, a single island connected to $M$ electronic reservoirs at $M$ QPCs is described by the multi-channel Kondo model, Eq.~\eqref{eq:MCKmodel} (with $\delta=0$)\cite{Matveev1991,Matveev1995}. In contrast to the small quantum dot implementation\cite{Glazman_Resonant_1988,ng_onsite_1988}, every electronic channel connected to the metallic island straightforwardly constitutes one additional Kondo channel\cite{Furusaki1995b}.

The measured charge-Kondo circuit is shown in Fig.~\ref{fig1}c. 
A micrometre-scale metallic island (colored gold) is thermally diffused into a (Al)GaAs semiconductor heterostructure to make contact with a buried two-dimensional electron gas (2DEG, dark gray).
The two consecutive charge states of the island forming $\boldsymbol{S}_\mathrm{K}$ can be tuned to degeneracy using the plunger gate voltage $V_\mathrm{pl}$ (top). The detuning $\delta V_\mathrm{pl}$ lifts the degeneracy of the charge states, equivalent to the Zeeman splitting when a magnetic field is applied to a real spin. 
The island charging energy $E_\mathrm{C}=e^2/2C\simeq 400$\,mK (with $C$ the island's geometrical capacitance, see Methods) sets an upper bound for the thermal energy $k_\mathrm{B}T$ where Kondo physics can develop.
Three Kondo channels can be individually connected using three QPCs, separately formed by applying a negative voltage to the corresponding capacitively coupled split gate (blue). 
Crosstalk between QPCs is reduced to a few percent by a grounded screening gate (light gray, below blue QPC gates). The transmission probability $\tau_i$ across QPC $i$ (which controls $J_i$) is measured in situations where the Kondo renormalization is minimal (Methods). 
Note that the real electron spin degeneracy is here lifted by the application of a large magnetic field $B\simeq5.3$\,T. This coincides with the integer quantum Hall regime at filling factor $\nu=2$, where the current propagates along chiral edge channels (white lines with arrows).

Monitoring the mean number of electrons on the island $\langle N \rangle$ is crucial for determining the impurity entropy via a Maxwell relation (discussed further below). 
We do this by measuring the conductance $G_\mathrm{sens}$ of an adjacent, capacitively coupled QPC (green in Fig.~\ref{fig1}c), which acts as a non-invasive charge sensor\cite{Field1993}. 
$G_\mathrm{sens}$ depends on $V_\mathrm{pl}$ and has period $\Delta$, see lower panel of Fig.~\ref{fig1}d. 
Relative to the charge degeneracy point defined as $\delta V_\mathrm{pl}=0$, we estimate the ``excess'' island charge $\langle \delta N\rangle\equiv \langle N\rangle_{\delta V_\mathrm{pl}}-\langle N\rangle_0$ at finite $\delta V_\mathrm{pl}$ from the conductance change of the charge sensor $\delta G_\mathrm{sens}=G_\mathrm{sens}(\delta V_\mathrm{pl})-G_\mathrm{sens}(0)$, via:
\begin{equation}
    \langle\delta N\rangle=\frac{\delta G_\mathrm{sens}}{\delta G_\mathrm{sens}^{1e}}+\frac{\delta V_\mathrm{pl}}{\Delta}\;, \label{eq:N(Gsens)}
\end{equation}
where $\delta G_\mathrm{sens}^{1e}$ is the change in sensor conductance when one electron is added (see Fig.~\ref{fig1}d).

\section*{Transport characterisation}

\noindent Before turning to entropy measurements, we fully characterize our device in terms of transport measurements. This serves two purposes. First, the analysis of conductance signatures will allow us to demonstrate the emergence of critical Kondo physics and quantitatively validate our experimental setup as an accurate quantum simulator for the multi-channel Kondo model, Eq.~\eqref{eq:MCKmodel}, by comparing to exact theoretical predictions. Second, for each configuration of the device, we will determine the Kondo temperature $T_\mathrm{K}$, an emergent scale controlling the onset of quantum criticality. 
Near the Kondo critical points at low temperatures $k_\mathrm{B}T\ll E_\mathrm{C}$, we expect \textit{universal scaling} of all physical quantities in terms of the rescaled parameter $T/T_\mathrm{K}$. 
This will be used later to establish universality in our entropy measurements.

We await 2CK quantum criticality\cite{Matveev1995, Furusaki1995b, Iftikhar2015} when two channels are coupled equally to the island, $\tau_1=\tau_2$ (and $\tau_3=0$). For three connected channels, 3CK criticality arises\cite{Iftikhar2018} when $\tau_1=\tau_2=\tau_3$. In both cases the system must be tuned to charge degeneracy, $\delta V_\mathrm{pl}=0$. The Kondo temperature $T_\mathrm{K}$ then depends on $\tau$ (which can be adjusted \textit{in-situ} and tuned to be equal for all channels) and $E_C$ (fixed by device geometry). 

The large $T/T_\mathrm{K}$ limit corresponds to the asymptotic-freedom regime where the Kondo effect is negligible. The island charge pseudospin is essentially free in this limit and the conductance is comparatively low. The exotic Kondo critical states of interest here are fully developed in the opposite limit of low $T/T_\mathrm{K}$ where the island pseudospin is partially screened and conductance is strongly Kondo-renormalized. 
In practice, at finite but low $k_\mathrm{B}T\ll E_C$ we achieve large $T/T_\mathrm{K}$ by tuning to weak bare coupling ($\tau\ll1$), whereas a stronger bare coupling ($1-\tau\ll1$ at 2CK but $\tau\sim0.8$ at 3CK) corresponds to small $T/T_\mathrm{K}$. Although quantum fluctuations extend to multiple island charge states\cite{Jezouin2016} near perfect transmission as $\tau\to1$, the low-temperature physics remains controlled by the multi-channel Kondo critical point\cite{Matveev1995,Furusaki1995b,Mitchell2016}, as confirmed experimentally in Ref.~\citenum{Iftikhar2018}.

\begin{figure}[ht]
        \centering
        \includegraphics[width=85mm]{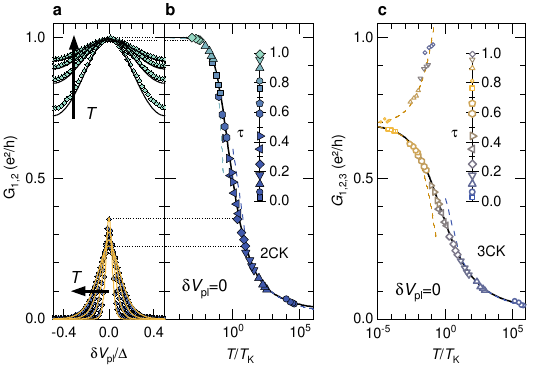}
        \caption{
        \footnotesize
        \textbf{Conductance benchmarks and Kondo criticality.}
             (a) Conductance vs plunger gate voltage at different temperatures for a 2CK circuit tuned either in the regime of weak coupling ($\tau\simeq0.25$, blue symbols) or strong coupling ($\tau\simeq0.993$, green symbols). 
             Lines represents analytical predictions in the corresponding limits. 
             (b,c) Universal Kondo conductance scaling at $\delta V_\mathrm{pl}=0$ vs $T/T_\mathrm{K}$ in two-channel (b) and three-channel (c) Kondo circuit configurations, for a broad range of couplings $\tau$. 
             Measurements shown as symbols (full for 2CK, open for 3CK) are confronted with NRG results (black solid lines) and asymptotic (colored dash) predictions. 
             The value of $T_\mathrm{K}$ for a given $\tau$ (data points displayed with the same symbol) is obtained as the best fit parameter. 
             The Kondo renormalization flow here shows in the temperature evolution for each configuration at fixed $\tau$ (or equivalently, fixed $T_\mathrm{K}$).
        } \label{fig2}
\end{figure}

We measure the renormalized conductance $G_i$ across the connected channels, as a function of $\tau$ and $\delta V_\mathrm{pl}$ at different electron temperatures $T$, following Ref.~\citenum{Iftikhar2018}. Our results are summarized in Fig.~\ref{fig2}.

Figure~\ref{fig2}(a) shows the channel conductances $G_{1,2}$ as a function of the detuning $\delta V_\mathrm{pl}$ from charge degeneracy, in a 2CK configuration ($G_1\simeq G_2$, $G_3=0$).
Colored symbols display measurements at several temperatures (from $T\simeq9.3$\,mK to $36.7$\,mK), for two representative values of the QPC transmissions. In the relatively weak-coupling regime ($\tau\equiv\tau_1\simeq\tau_2\simeq0.25$, blue symbols), the Kondo effect only weakly renormalizes the conductance, which accordingly exhibits narrow peaks. The line shapes here are well-described by those of an uncorrelated single-electron transistor (yellow lines, Methods). However, Kondo correlations are still evident in an effective reduction of the island charging energy\cite{Grabert1994} and a Kondo-enhanced peak height on decreasing temperature. 
In the opposite, strong-coupling regime ($\tau\simeq0.993$, green symbols) conductance data quantitatively follow analytical predictions\cite{Matveev1995,Furusaki1995b} for the limit of near-perfect transmission, specified completely without fit parameters in terms of the experimentally-characterized quantities $1-\tau_i\ll1$, $\delta V_\mathrm{pl}/\Delta$ and $k_\mathrm{B}T/E_\mathrm{C}\ll1$ (black lines, Methods). 

In Fig.~\ref{fig2}b,c, the temperature evolution of the measured conductance peak height at the charge degeneracy point $G_\mathrm{i}(\delta V_\mathrm{pl}=0)$ (symbols) is confronted with the universal conductance scaling in $T/T_\mathrm{K}$ predicted by the numerical renormalization group (NRG)\cite{Wilson1975} for the 2CK and 3CK models described by Eq.~\eqref{eq:MCKmodel} (black lines, Methods). 
Analytic asymptotic behaviors are displayed as colored dashed lines (Methods).
Data points represented by the same symbol correspond to the same value of $\tau$, measured at four different $T$. 
The corresponding Kondo temperature $T_\mathrm{K}(\tau)$ here is a fit parameter, separately adjusted for each $\tau$ (Methods). 
The data-theory agreement in Fig.~\ref{fig2}b,c should therefore be appreciated on the evolution with $T/T_\mathrm{K}$ (the slope) for any group of four identical symbols. 
At the Kondo critical fixed point itself (formally $T/T_\mathrm{K}\to0$) the conductance is predicted to be $G_\mathrm{2CK}=e^2/h$ for 2CK\cite{Furusaki1995b} and 
$G_\mathrm{3CK}=2\sin^2(\pi/5)e^2/h\simeq0.69e^2/h$ for 3CK\cite{Yi1998}, both of which are attained to high precision in the experiment. 
The excellent agreement between experiment and theory confirms that the investigated circuit matches the theoretical model, and realizes multi-channel Kondo criticality.

Although NRG results for $G>G_\mathrm{3CK}$ cannot be obtained in the 3CK setting, recent functional RG results\cite{paris_universal_2026} can access this regime.
In practice we estimate $T_\mathrm{K}$ here simply by fitting to the asymptotic power law scaling\cite{Affleck1993,Cox1998,Iftikhar2018} $G_{1,2,3}-G_\mathrm{3CK}\propto(T/T_\mathrm{K})^{2/5}$.
Symbol size indicates whether the data-theory comparison was performed against NRG predictions (large symbols) or asymptotic power laws (smaller symbols). 
For each setting, the extracted $T_\mathrm{K}(\tau)$ (shown in Methods) enables us to identify the quantity $T/T_\mathrm{K}$, which parameterizes the flow along the RG trajectory\cite{Wilson1975}. In turn, this affords a parameter-free analysis of the universal scaling of other physical quantities, such as the thermodynamic entropy considered next.

\section*{Impurity entropy via Maxwell relation}
    
\noindent We extract the entropy using the thermodynamic Maxwell relation\cite{child_entropy_2022,adam_entropy_2025} $\partial S/\partial \mu_\mathrm{e}=\partial\langle N\rangle/\partial T$. 
Variations in the number $\langle N\rangle$ of electrons on the island are determined using Eq.~\eqref{eq:N(Gsens)}, as a function of their electrochemical potential $\mu_\mathrm{e}$, itself controlled by the plunger gate voltage through $\delta \mu_\mathrm{e}=2 E_\mathrm{C} \delta V_\mathrm{pl}/\Delta$  (Methods). 
Integrating this relation from a given detuning $\Delta E$ up to the charge degeneracy point at $\delta \mu_\mathrm{e}=0$ gives the entropy \textit{difference} $\Delta S$ of the system between these different gate-voltage configurations: 
\begin{equation}
      \Delta S\equiv S_\mathrm{imp}(0)-S_\mathrm{imp}(\Delta E)=\int_{\Delta E}^{0}\frac{\partial\langle N \rangle}{\partial T}\Big{|}_{\delta V_\mathrm{pl}}\mathrm{d}\delta\mu_\mathrm{e}\;.
       \label{eq:DeltaS}
\end{equation}
The point of \textit{maximum} detuning, halfway between consecutive charge-degeneracy points, corresponds to $\Delta E=E_\mathrm{C}$ or equivalently $|\delta V_\mathrm{pl}|=\Delta/2$ (see Fig.~\ref{fig1}d). If the island charge is frozen at this point then its contribution to the entropy is negligible and $S_\mathrm{imp}(\Delta E)\simeq 0$. In this case $\Delta S \simeq S_\mathrm{imp}(0)$ directly gives the impurity entropy at the charge degeneracy point where Kondo criticality arises. This is guaranteed in the low-temperature limit $T\to 0$, as well as in the weak-coupling limit $\tau \to 0$. At finite temperatures and strong coupling, we expect a finite contribution to the entropy from $S_\mathrm{imp}(\Delta E)$ even at maximum detuning (Methods), which must be taken into account when interpreting these data. 

\begin{figure}[ht]
        \centering
        \includegraphics[width=84mm]{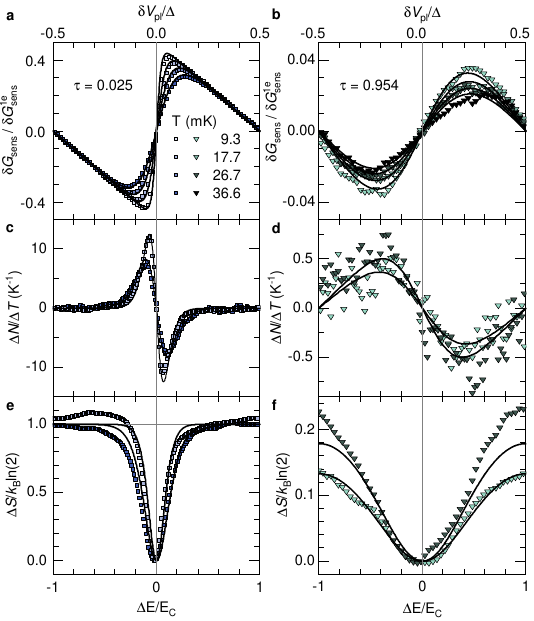}
        \caption{
        \footnotesize
        \textbf{Entropy measurement.} 
        Determination of the entropy for a device tuned to 2CK by measurement of the charge $N$ along sweeps in $V_\mathrm{pl}$ at different temperatures, for either weak (a,c,e) or strong (b,d,f) coupling $\tau$. Symbols with different shades correspond to different temperatures. Predictions (Methods) are displayed as continuous black lines. 
        (a,b) Charge measurements vs gate voltage ($\Delta E=2E_\mathrm{C}\delta V_\mathrm{pl}/\Delta$). 
        (c,d) Finite-difference approximation to the temperature derivative. 
        (e,f) Extracted impurity entropy difference.
        } \label{fig3}
\end{figure}

Figure~\ref{fig3} illustrates the procedure for 2CK configurations in the regimes of weak and strong couplings. 
Panels (a,b) show the charge-sensor measurement $\delta G_\mathrm{sens}/\delta G_\mathrm{sens}^{1e}$ vs gate voltage detuning $\delta V_\mathrm{pl}/\Delta$ at different temperatures $T$. 
The discrete derivatives $\Delta N/\Delta T$ displayed in (c,d) are performed with a large enough $\Delta T$ (between 10 and 20\,mK) to ensure sufficient signal while maintaining accuracy (Methods). 
This is then integrated from a given detuning $\Delta E$ up to the charge degeneracy point to give the entropy difference $\Delta S$, which is plotted in (e,f).

In the very weak-coupling regime ($\tau\simeq0.025$ in Fig.~\ref{fig3}a,c,e), the island charge pseudospin behaves as a free spin-$\tfrac{1}{2}$ impurity with an effective Zeeman splitting $\Delta E$. 
The ratio of the populations for the two charge states is therefore given by the Boltzmann factor $\exp(-\Delta E/k_\mathrm{B}T)$, which straightforwardly gives the predictions displayed as solid lines (Methods). The resulting $\Delta S$ displays a plateau at $k_\mathrm{B}\ln2$ at large detuning, corresponding to the entropy difference between the two configurations: a degenerate free spin at $\Delta E=0$, and a frozen impurity at $\Delta E=E_\mathrm{C}$. This validates our entropy extraction methodology at an accuracy of $\sim0.1\,k_\mathrm{B}\ln2$.
Note that the experimental accuracy is mainly limited by non-instrumental effects, such as residual crosstalks or localized charges (Methods).

In the strong-coupling regime ($\tau\simeq0.954$ in Fig.~\ref{fig3}b,d,f), the entropy profile strongly differs from the free-impurity prediction: it is broader and the maximum entropy change is reduced substantially below $k_\mathrm{B}\ln2$.  
In this regime the island charge remains incompletely frozen even at maximum detuning $\Delta E=E_\mathrm{C}$ (at $\delta V_\mathrm{pl}=\Delta/2$) and for our lowest $T\simeq 9$\,mK. 
This behavior (induced by the increasing quantum fluctuations at higher $\tau$\cite{Jezouin2016}) is directly evidenced by the non-zero conductance across the device (see insets in Fig.~\ref{fig4} and Methods).
Therefore $S_\mathrm{imp}(E_\mathrm{C})$ does not vanish, which reduces the measured $\Delta S$ with respect to the Kondo entropy $S_\mathrm{imp}(0)$ at large $\tau$. 
Nevertheless, the observed charge and entropy changes can still be predicted in the large-$\tau$ limit (using the same theory\cite{Matveev1995,han_fractional_2022} employed to describe the conductance in Fig.~\ref{fig2}a) and these are shown as the continuous lines in Fig.~\ref{fig3}b,d,f. Small deviations from the pristine $\tau\to1$ limit are accounted for here by an interaction-renormalized effective transmission $\tau=0.966$ (see Methods for a comparison using the bare $\tau=0.954$).
The experimental methodology is thereby established in both the weak and strong coupling limits. 

\section*{Impurity entropy renormalization flow}

\begin{figure}[hbt]
    \centering
    \includegraphics[width=85mm]{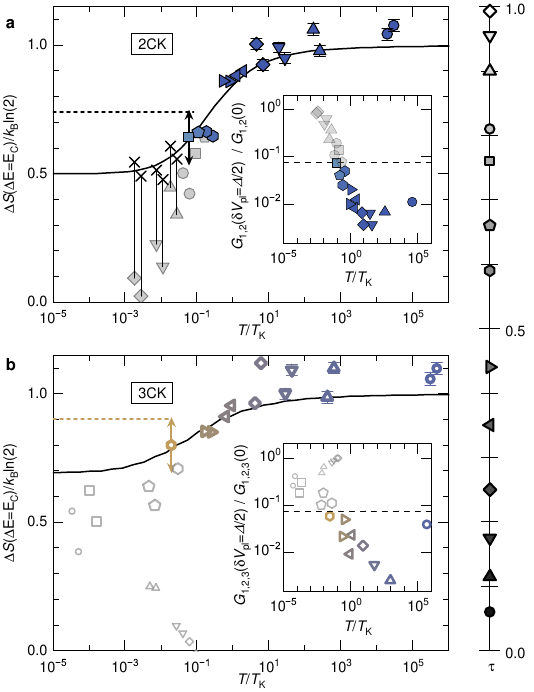}
    \caption{
    \footnotesize
    \textbf{Entropy renormalization flow} $\Delta S$ vs $T/T_{\mathrm{K}}$ for 2CK (a) and 3CK (b).
    Universal NRG predictions for $S_\mathrm{imp}(0)$ at the critical point are shown as solid lines. 
    Colored symbols represent experimental data points where we expect the measured $\Delta S$ to closely approximate the Kondo impurity entropy $S_\mathrm{imp}(0)$.
    These data are selected using the small conductance criterion $G_i(\Delta E=E_\mathrm{C})/G_i(0)<0.075$ which indicates a near-frozen impurity at maximum detuning, and hence a negligible contribution from $S_\mathrm{imp}(E_\mathrm{C})$, see insets.
    The horizontal dashed line in each main panel represents the associated upper bound for the Kondo critical entropy, taking into account the $\Delta S$ uncertainty estimated as $\pm 0.1\,k_\mathrm{B}\ln2$ (denoted $\updownarrow$). The smaller instrumental standard error is displayed as error bars when bigger than symbols.
    By contrast, the grey points correspond to cases where the conductance is appreciable even at maximum detuning, and the contribution $S_\mathrm{imp}(E_\mathrm{C})$ cannot be neglected. For 2CK, we also plot an estimate for $S_\mathrm{imp}(0)\simeq \Delta S + S_\mathrm{imp}^\mathrm{thy}(E_\mathrm{C})$ as the cross points for the largest three $\tau$ settings, using the theoretically-predicted entropy contribution at maximum detuning, $S_\mathrm{imp}^\mathrm{thy}(E_\mathrm{C}).$
    Symbols shape and size match those for identical configurations in Fig.~\ref{fig2}.
    }\label{fig4}
\end{figure}

The entropy difference $\Delta S$ between the charge degeneracy point where Kondo criticality occurs ($\delta V_\mathrm{pl}=0$) and maximum gate-voltage detuning ($\delta V_\mathrm{pl}=\Delta/2$) is obtained via Eq.~\eqref{eq:DeltaS} for different QPC transmissions $\tau$ and temperatures $T$. The dependence of the Kondo temperature $T_\mathrm{K}$ on $\tau$ (different for 2CK and 3CK, see Ext.\ Fig.~\ref{ExtFig4}) was separately extracted from conductance data at $\delta V_\mathrm{pl}=0$, allowing us to present our results for $\Delta S$ in terms of the RG flow parameter $T/T_\mathrm{K}$. 
We do this in Fig.~\ref{fig4} for both 2CK (a) and 3CK (b) circuit configurations (see Ext.\ Fig.~\ref{ExtFig7} for 1CK).
Experimental results (symbols) are confronted with universal NRG predictions for $S_\mathrm{imp}(0)$ at the 2CK and 3CK critical points (black solid lines).

Colored symbols correspond to configurations where we expect $\Delta S \approx S_\mathrm{imp}(0)$. This is based on an independent measurement of the conductance at $\delta V_\mathrm{pl}=\Delta/2$: when the island charge is frozen at maximum detuning, it contributes negligible entropy $S_\mathrm{imp}(\Delta/2)\simeq 0$ but this also implies that the conductance through the island vanishes. We therefore only interpret $\Delta S$ as approximating the desired $S_{\rm imp}(0)$ when the conductance at maximum detuning is below the threshold value indicated by the horizontal dashed line in the insets (Methods). Taking only the ``trusted'' colored data points, we still observe a clear decrease in the measured entropy around $T/T_\mathrm{K}\sim1$, precisely where the Kondo effect sets in. Moreover, the evolution of the entropy upon decreasing $T/T_\mathrm{K}$ closely matches the theoretical curves.
This experimentally establishes the predicted universal 2CK and 3CK entropy renormalization flows over a broad span of $T/T_\mathrm{K}$ $(\gtrsim0.1)$.
Although not extending to a full saturation at $\Delta S=k_\mathrm{B}\ln\sqrt{2}$ for 2CK or $k_\mathrm{B}\ln\phi$ for 3CK, these data provide an \textit{upper bound} on the critical value of $S_\mathrm{imp}(0)$. 
With a minimal assumption of entropy monotonicity, we find  $S_\mathrm{imp}(0)<0.74\,(0.90)\,k_\mathrm{B}\ln2$ 
for the 2CK (3CK) ground states, indicated by horizontal dashed lines in the main panels, which is obtained by adding the estimated $0.1\,k_\mathrm{B}\ln2$ uncertainty to the lowest measured point for which we are confident that $\Delta S\approx S_\mathrm{imp}(0)$.
Note that, consistent with the vanishing Kondo entropy predicted at 1CK, a markedly smaller upper bound $S_\mathrm{imp}(0)<0.23\,k_\mathrm{B}\ln2$ can be obtained along the same lines in the one-channel case (Methods).

Grey points are for settings where $S_\mathrm{imp}(\Delta/2)$ is non-negligible (diagnosed by appreciable conductance through the island even at maximum detuning, see insets), and hence where $\Delta S<S_\mathrm{imp}(0)$ falls below the displayed Kondo entropy predictions.
The behaviors for 2CK and 3CK are qualitatively different. For 2CK, the measured $\Delta S$ decreases steadily to 0 as $T/T_\mathrm{K}$ is reduced. This is because the lowest $T/T_\mathrm{K}$ arises as $\tau\to 1$ in the 2CK setting, and this is precisely the regime where strong quantum fluctuations persist even at maximum detuning, strongly suppressing the measured $\Delta S$.   
By contrast, for 3CK the measured $\Delta S$ remains finite at the lowest $T/T_\mathrm{K}$. This is a consequence of the non-monotonic dependence of $T_\mathrm{K}$ on $\tau$ (see Ext.\ Fig.~\ref{ExtFig4}). In this setting, the lowest $T/T_\mathrm{K}$ are achieved at $\tau\approx0.8$, for which moderate fluctuations at maximum detuning only partially reduce $\Delta S$ below $S_{\rm imp}(0)$. Full suppression of $\Delta S$ all the way to zero only occurs as $\tau \to 1$, as shown by the small grey symbols in Fig.~\ref{fig4}b, corresponding to intermediate $T_{\rm K}$. 

In the 2CK setting in the strong coupling regime, theoretical predictions for the Coulomb valley entropy $S^\mathrm{thy}_\mathrm{imp}(\delta V_\mathrm{pl}=\Delta/2)$ can be obtained from Refs.~\citenum{Matveev1995,han_fractional_2022} (Methods). 
This provides a means of estimating the critical Kondo impurity entropy via $S_\mathrm{imp}(0)\simeq\Delta S+S^\mathrm{thy}_\mathrm{imp}(\delta V_\mathrm{pl}=\Delta/2)$, where $\Delta S$ is measured. 
This correction is shown in Fig.~\ref{fig4}a as the cross points for the data at largest $\tau$, and correctly captures the saturation of $S_\mathrm{imp}(0)$ towards the finite, fractional value of $k_\mathrm{B}\ln\sqrt{2}$.
However, this is shown only as a self-consistent check; our key result for the upper bound on the critical Kondo entropy is obtained directly from the experimental data and does not rely on model predictions.

Having obtained conservative upper bounds on the 2CK and 3CK ground state entropies by examining the universal entropy flows to their respective critical points, in the next section we obtain complementary \textit{lower bounds} by analyzing the entropy flows \textit{away} from the critical points, induced by small gate detunings. These results must be taken together to establish fractional Kondo critical entropies.

\section*{Universal Fermi-liquid crossover}

\noindent At $\delta V_\mathrm{pl}=0$, the physics is entirely governed by the 2CK or 3CK critical fixed points at low temperatures $T\ll T_\mathrm{K}$. In particular, the predicted $T=0$ residual entropy $S_\mathrm{imp}$ is finite -- symptomatic of the frustrated Kondo ground state of the system. Small gate voltage perturbations relieve this frustration, and result in the entropy being quenched as the system flows away from the critical point, towards a regular Fermi-liquid ground state. This crossover, characterized by the gate-voltage-induced scale $T_\mathrm{co}$, is a universal single-parameter scaling function of $T_\mathrm{co}/T$ provided (i) critical physics is well-developed ($T\ll T_\mathrm{K}$); and (ii) the perturbation is small ($T_\mathrm{co}\ll T_\mathrm{K}$). 
Transport signatures of this Fermi-liquid crossover have been established in both 2CK\cite{mitchell2012universal,Keller2015,Mitchell2016,Iftikhar2018} and 3CK\cite{Iftikhar2018} settings, but the entropy should also exhibit universality in this regime. 
We seek to demonstrate such behavior, and with it to provide experimental lower bounds for the ground state values of $S_\mathrm{imp}$ at the Kondo critical points.

 \begin{figure}[ht]
    \centering
    \includegraphics[width=85mm]{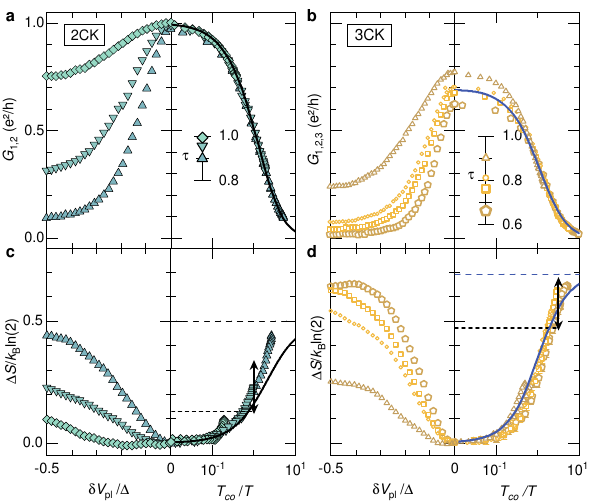}
    \caption{
    \footnotesize
    \textbf{Crossover flow from quantum criticality} at 2CK (a,c) and 3CK (b,d).
    \textbf{a,b} Conductance measurements at $T=9.3$\,mK are plotted as symbols versus $\delta V_\mathrm{pl}/\Delta$ (left panel) or versus $T_\mathrm{co}/T$ (right panel) for different $\tau$, chosen such that $T/T_\mathrm{K}<0.02$ to ensure well-developed criticality at charge degeneracy. The prefactor $\lambda_\mathrm{co}$ in Eq.~\eqref{eq:Tco_main} is adjusted for each $\tau$ to match universal crossover predictions for the conductance (black solid lines). 
    \textbf{c,d} Impurity entropy change $\Delta S$ along the Fermi-liquid crossover for the same settings vs the same $\delta V_\mathrm{pl}/\Delta$ (left) or $T_{\mathrm{co}}/T$ (right). 
    The $\Delta S$ universality here is observed without any adjustable parameter, since $\lambda_\mathrm{co}$ is separately extracted from the conductance evolution.
    Lower horizontal dashed lines represent the associated experimental lower bounds for the 2CK and 3CK critical entropies, taking into account an experimental uncertainty of $\pm 0.1\,k_\mathrm{B}\ln2$ ($\updownarrow$).
    Theory scaling curves are shown as solid lines. Upper dashed lines indicate the predicted ground-state Kondo fractional entropies. 
    }\label{fig5}
\end{figure} 

Figure~\ref{fig5}a,b shows conductance sweeps measured at 9.3\,mK for 2CK (a) and 3CK (b) configurations, as a function of gate-voltage detuning $\delta V_\mathrm{pl}$ (left). We choose $\tau$ values such that quantum criticality is rather well-developed at the charge degeneracy point $\delta V_\mathrm{pl}=0$
(in practice $T/T_{\mathrm{K}}<0.02$, corresponding to predicted impurity entropy deviations below $0.1\,k_\mathrm{B}\ln2$ from the Kondo ground states, see Methods).
Finite $\delta V_\mathrm{pl}$ induces a finite crossover scale $T_\mathrm{co}$, and we confirm universal scaling collapse of the conductance data (symbols) in terms of $T_\mathrm{co}/T$ in the right panels of Fig.~\ref{fig5}a,b compared with NRG predictions (solid lines). 
For 2CK, Ref.~\citenum{Furusaki1995b} gives an exact expression for $T_\mathrm{co}$ in terms of the experimentally-characterized parameter $\delta V_\mathrm{pl}/\Delta$,
\begin{equation}\label{eq:Tco_main}
  T_\mathrm{co}=\lambda_\mathrm{co}\sin^a(\pi\delta V_{\mathrm{pl}}/\Delta)\;,  
\end{equation}
with critical exponent $a=2$. The prefactor $\lambda_{\mathrm{co}}$, which depends on the tuning of $\tau$, is the only adjustable parameter for $ G(T_{\mathrm{co}}/T)$ to match the NRG predictions. For 3CK, an exact expression is not known, but Eq.~\eqref{eq:Tco_main} with $a=5/3$ was shown to be a reliable ansatz in the experiments of Ref.~\citenum{Iftikhar2018} (and correctly captures the known\cite{Affleck1993} low-detuning power law $\delta V_\mathrm{pl}^{5/3}$). 
Figure~\ref{fig5}c,d shows the corresponding entropy $\Delta S$ vs gate-voltage detuning $\delta V_\mathrm{pl}$ (left), and the universal scaling collapse of the data when plotted in terms of $T_\mathrm{co}/T$ (right, using $T_\mathrm{co}$ separately extracted from conductance). 
The observed collapse experimentally demonstrates the universality of the crossover for the entropy.
Although these data do not extend along the full crossover, the implications are: 
(i) The universality of the crossover in $T_\mathrm{co}/T$ for small detuning from degeneracy, observed on two observables, provides independent evidence not relying on the Kondo model that the system is deep in the quantum critical regime at the charge-degenerate critical point. 
Accordingly, the critical entropy $S_\mathrm{imp}(0)$ for these settings is near its low-temperature limit. 
This conclusion is further ascertained from the vicinity of the conductance to its low-temperature fixed point value (corresponding to $T/T_\mathrm{K}<0.02$).
(ii) The critical entropy is larger than $\Delta S$ measured at the highest $T_\mathrm{co}/T$ for which universal scaling is ascertained. This immediately follows from the fact that the universal $\Delta S(T_\mathrm{co}/T)$ crossovers are evolving from 0 (at $T_\mathrm{co}\simeq 0$) toward the critical entropy value as $T_\mathrm{co}/T$ increases (together with a minimal assumption of monotonicity).
In practice, we ascertain the crossover's universality by requiring that both entropy and conductance measurements vs $T_\mathrm{co}/T$ overlap for at least two different device settings corresponding to $T\ll T_\mathrm{K}$.
(iii) The 2CK and 3CK universal crossover curves are markedly different, indicating flows to different critical points.
With these implications, taking into account the $0.1\,k_\mathrm{B}\ln2$ experimental uncertainty, we obtain experimental lower bounds on the critical values of $S_\mathrm{imp}(0)$: $S_\mathrm{imp}\gtrsim0.13\,k_\mathrm{B}\ln2$ and $\gtrsim0.47\,k_\mathrm{B}\ln2$ for the 2CK and 3CK ground states (lower horizontal dashed lines in Fig.~\ref{fig5}c,d).

Our conclusions are further strengthened by comparing the $\Delta S$ experimental data with universal theoretical predictions (solid lines) when plotted vs $T_\mathrm{co}/T$. 
Theory lines are obtained in the limit of very small detuning and very small $T/T_\mathrm{K}$.
We observe, without any adjustable parameters, a data-theory agreement that remains within $\sim0.1\,k_\mathrm{B}\ln2$ and extends over the main part of the crossover curves, up to $T_\mathrm{co}/T\sim10$. Note that data-theory deviations increase for the largest detunings, where universality is eventually expected to break down.
Furthermore, the clear hierarchy observed between the Kondo entropy flows (Fig.~\ref{fig4} and Ext.\ Fig.~\ref{ExtFig7}) already suggests a higher, therefore non-zero, ground state entropy at 3CK and 2CK compared with 1CK.

\section*{Discussion}

\noindent With a tunable and fully characterized charge-Kondo circuit, we investigated the impurity entropy emerging in the two-channel and three-channel Kondo quantum-critical states. 
We employed a thermodynamic Maxwell relation to extract the entropy, which constitutes a direct, conjecture-free methodology. 

Our results for the entropy at charge degeneracy were found to match the predicted 2CK and 3CK universal flows over a large span of the explored $T/T_\mathrm{K}$ (see Methods for 1CK). Our analysis places an experimental \textit{upper} bound on the fractional entropy values associated with the frustrated Kondo ground states. 
Furthermore, the entropy crossovers on the scale of $T_\mathrm{co}$, induced by detuning away from charge degeneracy in well-developed 2CK and 3CK states, were found to obey universal scalings in terms of $T_\mathrm{co}/T$, again in specific agreement with theoretical predictions. The large explored range of $T_\mathrm{co}/T$ provides a \textit{lower} bound for the 2CK and 3CK ground state entropies. 

Taken together, our experimental results establish the finite, fractional nature of the ground-state entropy in two-channel and three-channel Kondo systems. Accounting for estimated experimental uncertainties, we find 
$[0.13,0.74]\,k_\mathrm{B}\ln 2$ for 2CK and $[0.47,0.9]\,k_\mathrm{B}\ln 2$ for 3CK, which are consistent with the predicted values\cite{affleck1991universal} of $S_\mathrm{imp}=k_\mathrm{B}\ln\sqrt{2}$ for 2CK, corresponding to a free Majorana zero mode, and $k_\mathrm{B}\ln\tfrac{1+\sqrt{5}}{2}\simeq0.69\,k_\mathrm{B}\ln 2$ for 3CK, corresponding to a free Fibonacci anyon.

In general, the coexistence of multiple types of Abelian anyons, such as in the Laughlin fractional quantum Hall states, could give rise to a fractional ground state entropy. 
However, in the present context of an emerging individual Kondo anyon, the observed fractional impurity entropy constitutes a direct fingerprint of its fundamental non-Abelian character.
An array of such localized anyons, each tied to a separate charge-Kondo impurity, has been proposed as a platform for measurement-based topological quantum computation, although the interconnection between impurities remains a practical challenge\cite{lopes2020anyons,Lotem2022}.
More broadly, the entropy characterisation protocol developed here could be extended to other tunable platforms hosting exotic quasiparticles and quantum criticality, including fractional quantum Hall systems\cite{sankar_measuring_2023} and strongly correlated nanostructures\cite{LeHur2004,Potok2007,Mebrahtu2013,SelaMZM2019,Nguyen2018seebeck2ICK,karki2023z}.

\vskip 1cm
{\Large\noindent\textbf{METHODS}}
\small

\noindent\textbf{Cryogenic setup.}
The sample was thermally anchored to the mixing chamber of a cryofree dilution refrigerator, enclosed in two shields at base temperature, and electrically connected by lines extensively filtered and thermalized (see Ref.~\citenum{Iftikhar2016} for details). 
Standard lock-in techniques at low-frequencies (below 200\,Hz) were used to extract the differential conductances across the charge-Kondo circuit and the charge sensor, using ac excitations of rms voltage below $k_\mathrm{B}T/e$.
Noise thermometry measurements were performed with home-made cryogenic amplifiers, around the 0.9\,MHz resonant frequency of a tank circuit\cite{jezouin_quantum_2013}.\\

\noindent\textbf{Nanofabrication.} 
The circuit is nanofabricated from a GaAs/AlGaAs heterostructure hosting, 90\,nm below the surface, a two-dimensional electron gas (2DEG) of electron density $n_e=2.6\times10^{11}\,\mathrm{cm}^{-2}$ and mobility $\mu=0.5\times10^6\,\mathrm{cm}^2/\mathrm{Vs}$.
The 2DEG mesa is first delimited by a 100\,nm deep wet etching using a solution of $\mathrm{H_3PO_4/H_2O_2/H_2O}$ (the 2DEG is etched away in lighter gray area of Fig.~\ref{fig1}c).
Low resistance ohmic contacts to the buried 2DEG are then realized by thermal annealing of a metallic stack Ni/Au/Ge/Ni/Au/Ni (10/10/90/20/170/40\,nm). 
Third, a first HfO$_2$ insulating and protective layer of nominal thickness 15\,nm, grown by ALD, covers the all device.
Fourth, a Ti(5\,nm)/Au(20\,nm) bilayer forms a bottom screening gate (appearing whiter in Fig.~\ref{fig1}c) located below and around most of the subsequently deposited QPC gates (except near the point contact), and also forms one gate controlling the sensor (colored green, pointing toward the island). 
Fifth, a second ALD-grown insulating layer of HfO$_2$ of nominal thickness 15\,nm covers the sample.
Sixth, 40\,nm of Al forms the QPC gates (blue), the sensor barring gate (green, closest to the island), and a grounded screening layer (light gray) above and around the previously deposited Ti/Au sensor gate (except near the end tip where the sensor point contact is located).
Seventh, a thick bilayer of Cr(5\,nm)/Au(400\,nm) makes contact on the outer part of the device with the previously deposited Al layer to provide robust bonding pads.\\

\noindent\textbf{Cross-talk.}
The capacitive coupling between island, and charge sensor is essential for charge detection.
However, the electrostatic coupling between each QPC gate and the other QPCs or the charge sensor, as well as between the two sensor gates and the QPCs, introduces unwanted complications.
In particular, tuning one QPC can thereby modify the tuning of the other ones and shift the sensor working point.
This effect is mitigated by the relatively important (several-micrometre) separation between gates controlling different circuit elements (see Fig.~\ref{fig1}c), and also by the implementation of bottom and top screening gates (see `Nanofabrication'). 
Residual cross-talks are systematically calibrated. 
Their relatively small values, found between 0.05\% and 2\%, allow us to perform high-precision compensation with straightforward linear corrections.
Nevertheless, we suspect that imperfect compensation toward the very sensitive charge sensor (e.g.\ due to a dependence of the cross-talk with device tuning) may contribute significantly to the observed $\pm0.1\,k_\mathrm{B}\ln2$ experimental uncertainty on entropy.\\

\noindent\textbf{Charging energy.}
A first estimate of the charging energy $E_\mathrm{C}=e^2/2C$ of the metallic island is obtained from Coulomb diamond measurements with two QPCs set to a weak, tunnel coupling.
Extended Fig.~\ref{ExtFig1} displays the color-coded differential conductance $dI/dV$ across the device as a function of dc bias $V_\mathrm{dc}$ and $V_\mathrm{pl}$.
The height of a diamond in voltage bias corresponds to $E_\mathrm{C}/2e$, yielding a charging energy $E_\mathrm{C}\simeq35\pm4\,\mu\text{eV}$, or approximately $k_\mathrm{B}\times400$\,mK.

\begin{figure}[t]
\setcounter{figure}{0}
\renewcommand{\figurename}{\textbf{Extended Figure}}
\renewcommand{\thefigure}{\textbf{\arabic{figure}}}
\centering\includegraphics [width=1\columnwidth]{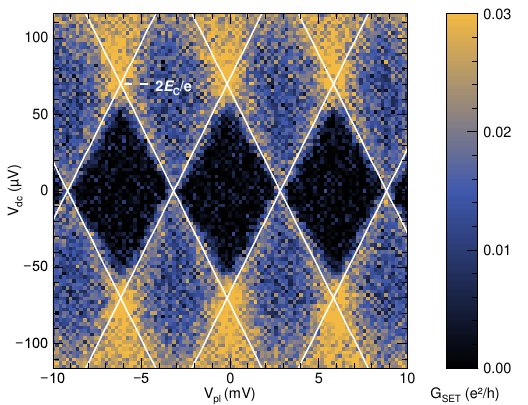}
\caption{
\footnotesize
\textbf{Coulomb diamond} characterisation of the charging energy.
The color-coded differential conductance $G_\mathrm{SET}$ across the device connected by two tunnel QPCs is displayed versus plunger gate voltage ($V_\mathrm{pl}$) and dc voltage ($V_\mathrm{dc})$.
The $E_\mathrm{C}=34.9\,\mu$eV prediction for the Coulomb diamond edges is shown as white straight lines. 
}
\label{ExtFig1}
\end{figure}

This estimate is confirmed and improved using the charge sensor.  
Specifically, we compare the sensor conductance change for one additional electron jumping in the tunnel-coupled island ($\delta G_\mathrm{sens}^{1e}$, see Fig.~\ref{fig1}d) to the signal change $\delta G_\mathrm{sens}^{\Delta V}$ caused by the known potential shift $\Delta V$ applied to the island by a large electrode when it is connected through a ballistic channel ($\tau=1$) such that charge quantization is fully quenched (otherwise the island's potential would not follow linearly the applied voltage). 
Within linear cross-talk compensations for the sensor QPC, the charging energy is simply given by:
\begin{equation}
E_\mathrm{C}=\frac{\delta G_\mathrm{sens}^{1e}}{\delta G_\mathrm{sens}^{\Delta V}} \times \frac{e \Delta V}{2}.
\end{equation}
This independent determination of the charging energy with the charge-Kondo circuit tuned in the opposite, ballistic limit gives $E_\mathrm{C}=34.9\pm1.0~\mu$eV, consistent with the standard Coulomb diamonds approach.

Note that a higher charging energy is particularly desirable. In the present Kondo entropy context, a larger $E_\mathrm{C}$ would allow us to directly observe the fractional entropy, by increasing the range of temperature and transmission for which the impurity is frozen in the Coulomb valley (where $\Delta S(E_\mathrm{C})\simeq S_\mathrm{imp}(0)$). 
In our GaAs platform, the challenge is to preserve good electrical contacts between the metallic island and the buried 2DEG, which is incompatible with present technologies to sub-micrometre island sizes. 
Promising alternatives are platforms based on InAs (of Fermi level pinning within conduction band) or on graphene (much closer to the surface), see Ref.~\citenum{sriram2025inas} for a first step with InAs.\\

\noindent\textbf{QPC tuning and characterisation.} 
The bare (unrenormalized by Kondo and charging effects) transmission probability $\tau_{1,2,3}$ across the single electronic channel connected across each of QPC$_{1,2,3}$ determines the corresponding Kondo coupling.
Importantly, the 2-3CK quantum-critical points are at precisely identical couplings for each connected channel, whatever the coupling strength.
It is therefore most crucial to achieve a precise symmetry. 
It was finely adjusted in the 3CK configurations, where $G_{1,2,3}$ can be individually measured, at the intermediate temperature $T=17.6\,\mathrm{mK}$ where possible temperature differences between contacts are less likely. 
Note that with only two connected channels, it is not possible to separately extract $G_1$ and $G_2$ from conductance measurements across the whole circuit.
The symmetry between QPC$_1$ and QPC$_2$ was preserved by compensating for the crosstalk induced by closing QPC$_3$.

\begin{figure}[t]
\renewcommand{\figurename}{\textbf{Extended Figure}}
\renewcommand{\thefigure}{\textbf{\arabic{figure}}}
\centering\includegraphics [width=1\columnwidth]{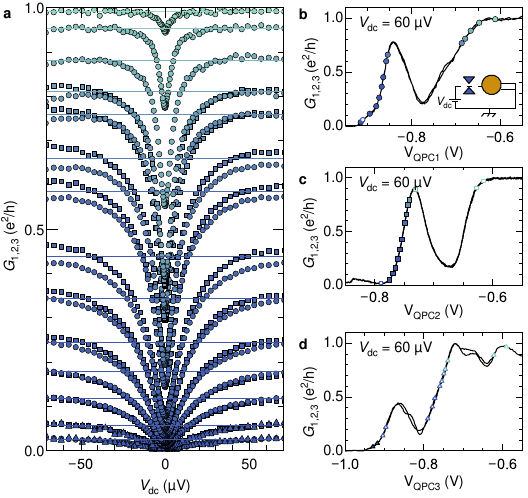}
\caption{
\footnotesize
\textbf{QPC tuning}.  
\textbf{a}, QPC differential conductance vs.~dc voltage measured at $T=9.3$\,mK in the dynamical Coulomb blockade configuration (see schematic in b) for the tunings shown as full symbols in b–d. 
Horizontal lines indicate the corresponding bare (unrenormalized) transmissions $\tau$, identical for all channels, extracted for fourteen implemented QPC tunings.  
\textbf{b–d}, Differential conductances $G_{1,2,3}$ measured under a relatively high bias $V_\mathrm{dc}=60\,\mu$eV versus the voltage applied to the respective QPC split gate. Continuous lines are bidirectional gate sweeps; symbols mark the implemented tunings. Filled symbols correspond to retained configurations, while open symbols indicate discarded ones based on anomalous or asymmetric energy dependence. 
}
\label{ExtFig2}
\end{figure}

A quantitative characterisation of the corresponding $\tau$ is also essential for direct comparison with theory, particularly in the near-ballistic limit $1-\tau \ll 1$. 
The bare transmissions $\tau_{1,2,3}$ of the QPCs were extracted from their conductances $G_{1,2,3}$ measured at relatively high voltage bias, while the other QPCs were tuned to the well-defined ballistic transmission plateau ($\tau=1$), as schematically illustrated in the inset of Ext.\ Fig.~\ref{ExtFig2}b. 
In this configuration, the island's charge quantization is lifted but $G_{1,2,3}$ is suppressed at low bias and temperature by the so-called dynamical Coulomb blockade (DCB), see Ext.\ Fig.~\ref{ExtFig2}a. 
In practice, we have selected fourteen different symmetric tunings, identical for each QPCs and spanning the whole range of transmission probabilities. The bare transmissions $\tau_{1,2,3}$ were then obtained as the high-bias averages of $G_{1,2,3}$ (horizontal solid lines).  Their position within the corresponding QPC conductance vs gate voltage sweep is displayed by symbols in Ext.\ Fig.~\ref{ExtFig2}b-d
However, some settings exhibited anomalous energy dependences, likely due to localized defects near the constriction, preventing reliable extraction of $\tau$. To exclude such cases, canonical settings were selected using two criteria: (i) the energy dependence of all three QPCs in a symmetric configuration had to be consistent; and (ii) the conductance energy dependence had to remain symmetric. Only settings fulfilling both criteria were retained (filled symbols in Ext.\ Fig.~\ref{ExtFig2}), while discarded ones are shown as open symbols in panels b–d.\\

\begin{figure}[t]
\renewcommand{\figurename}{\textbf{Extended Figure}}
\renewcommand{\thefigure}{\textbf{\arabic{figure}}}
\centering\includegraphics [width=1\columnwidth]{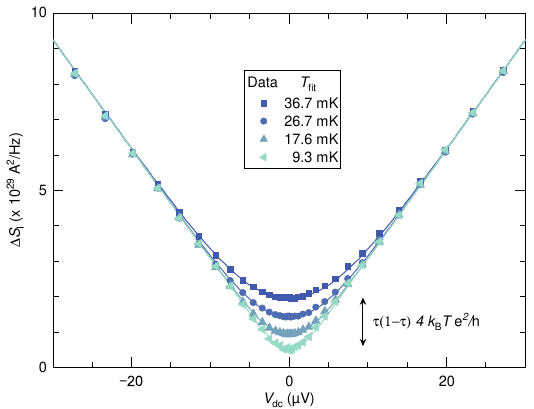}
\caption{
\footnotesize
\textbf{Shot-noise thermometry.} 
The excess spectral density $\Delta S_\mathrm{I}= S_\mathrm{I}(V_\mathrm{dc})-S_\mathrm{I}(0)$ of the current fluctuations (symbols) is plotted versus the bias $V_\mathrm{dc}$ for different temperatures. 
Fitting with Eq.~\eqref{eq:shotnoise} (lines) enables the extraction of the electron temperature $T$ indicated in the symbols legend.
Displayed data and predictions are shifted vertically by the corresponding thermal shot noise contribution $\tau(1-\tau)4k_\mathrm{B}Te^2/h$.
}
\label{ExtFig3}
\end{figure}

\noindent\textbf{Electronic temperature}.
The electronic temperature $T$ was determined from standard shot-noise thermometry.
We used for this purpose the QPC closest to the sensor, and opened the sensor barring gate (green gate closest to the island in Fig.~\ref{fig1}c) to bypass the island (whose heating at finite bias would otherwise perturb the thermometry).
The corresponding excess current noise data $\Delta S_\mathrm{I}\equiv S_\mathrm{I}(V_\mathrm{dc}) - S_\mathrm{I}(0)$ is shown as symbols vs QPC voltage bias in Ext.\ Fig.~\ref{ExtFig3}.
The electronic temperature is obtained by fitting these data with the expression:
\begin{equation}
\Delta S_I=\frac{2e^2}{h}\tau(1-\tau)eV_\mathrm{dc}\left[ \coth\left(\frac{eV_\mathrm{dc}}{2k_\mathrm{B}T}\right)-\frac{2k_\mathrm{B}T}{eV_\mathrm{dc}}\right].
\label{eq:shotnoise}
\end{equation}
Note that the extracted $T$ was cross-validated by Johnson–Nyquist thermometry on the large ohmic contacts.
All data presented in the main manuscript were acquired at the electronic temperatures of
$T=9.3\pm0.1$\,mK, $17.6\pm0.1$\,mK, $26.7\pm0.3$\,mK and $36.7\pm0.3$\,mK.\\

\noindent\textbf{Reproducibility criteria.}
Charge sensor data were collected over repeated plunger gate sweeps, each spanning multiple gate voltage periods. Within the window $\lvert \delta V_\mathrm{pl} \rvert < \Delta/2$, the displayed data correspond to an ensemble average of at least 20 individual, single-period sweeps. To ensure the robustness of the extracted signal and eliminate the influence of transient noise or charge instabilities, a two-step check based on data reproducibility is implemented.
First, we test the data reproducibility on two quantities, namely (i) the integral of the charge sensor signal over one full period and (ii) the integral of the absolute difference between the forward and reverse sweep over the same period. 
Single-period sweeps for which one of these integrated values deviates from the ensemble mean by more than three standard deviations were automatically discarded.
Note that the probability to find such data points in a Gaussian distribution is of 0.3\%.
Second, we test the reproducibility of individual data points on the retained single-period sweeps.
For each value of $\delta V_\mathrm{pl}$, we check whether the charge sensor signal is less than three standard deviations away from the ensemble mean at this specific $\delta V_\mathrm{pl}$. 
If it is further away, this individual data point is automatically discarded from the averaged data.
Overall, these reproducibility criteria typically remove less than 10\% of the data.\\

\noindent\textbf{$\Delta S$ step-by-step.}
After applying the reproducibility criteria detailed above, the multiple sweeps are averaged. With the averaged data, the charge-sensor signal is calibrated (for every temperature) and then converted into the mean electron number in the island. Repeating the process at different temperatures allows us to compute $\Delta N/\Delta T$, which is integrated to obtain $\Delta S$.\\
In more detail, for each temperature the charge-sensor response to a single electron ($\delta G^{\mathrm{sens}}_{1e}$) is calibrated using the configuration with the weakest impurity--bath coupling, where the island's charge remains quantized.
By comparison with measurements performed in the fully open regime under a known island dc bias, we verify that the sensor sensitivity is essentially independent, up to a few percent, of the device setting.\\
With $\delta G^{\mathrm{sens}}_{1e}$ calibrated, the variation of the mean electron number $\delta N$ is obtained using Eq.~(\ref{eq:N(Gsens)}). The temperature derivative is then evaluated using a discrete temperature differentiation:
\begin{equation}
\frac{\Delta N}{\Delta T}(\delta V_{\mathrm{pl}})=
\frac{\langle \delta N(\delta V_{\mathrm{pl}},T_j)\rangle - \langle \delta N(\delta V_{\mathrm{pl}},T_i)\rangle}{T_j-T_i}.
\end{equation}
Closely spaced temperature pairs ($15~\mathrm{mK}<|T_j-T_i|<20~\mathrm{mK}$) are used to ensure a sufficiently approximation of the derivative, while preserving a reasonable signal-to-noise ratio. 
In practice, we used the pairs 26.7--9.3\,mK and 36.6--17.7\,mK. Each $\Delta N/\Delta T$ curve shown in Fig.~3(c,d) corresponds to one such pair.\\
Finally, the entropy difference $\Delta S$ is obtained by integrating $\Delta N/\Delta T$ with respect to the plunger-gate voltage using the Maxwell relation Eq.~(\ref{eq:DeltaS}).
This yields the entropy difference between the charge-degeneracy point ($\delta V_\mathrm{pl}=0$) and the chosen detuning.\\

\noindent\textbf{Experimental uncertainty.}
A first source of uncertainty is the electrical (amplifier and thermal) noise. 
The resulting standard error on the QPC and the sensor conductances is typically found within $(1{-}2)\times10^{-3}\,e^2/h$ and $(1{-}2)\times10^{-4}\,e^2/h$, respectively, depending on temperature. 
These values correspond to an uncertainty of less than 0.2\% on the QPC transmissions and of about 0.5\% of $\delta G_\mathrm{sens}^{1e}$.
For the charge and the entropy, an additional source of uncertainty is the precision of the one electron sensor signal calibration $\delta G_\mathrm{sens}^{1e}$.
It lies between 0.5\% and 0.8\% across all measured temperatures. 
Furthermore, a specific source of uncertainty for the entropy is the experimental resolution on the electronic temperature, typically found within $\pm 1\%$.
Altogether, we expect from these individual error sources that the entropy should be resolved at an estimated accuracy similar or better than $0.03\,k_\mathrm{B}\ln2$. This uncertainty, computed for every settings, is displayed as error bars in Fig.~\ref{fig4} if larger than symbols. However, in  practice, we observe a larger scatter of the entropy data.
This behavior may originate from a combination of imperfections, such as slight changes of the cross-talk corrections with the tuning of the device, or the influence of charged impurities, sufficiently weak or reproducible to evade automatic detection from our data reproducibility criteria.
A more direct estimate of the experimental entropy uncertainty, directly based on the actual scatter of entropy data points in Fig.~\ref{fig3}e and Fig.~\ref{fig4}, is $\lesssim0.1\,k_\mathrm{B}\ln2$.\\

\begin{figure}[t]
\renewcommand{\figurename}{\textbf{Extended Figure}}
\renewcommand{\thefigure}{\textbf{\arabic{figure}}}
\centering\includegraphics [width=89mm]{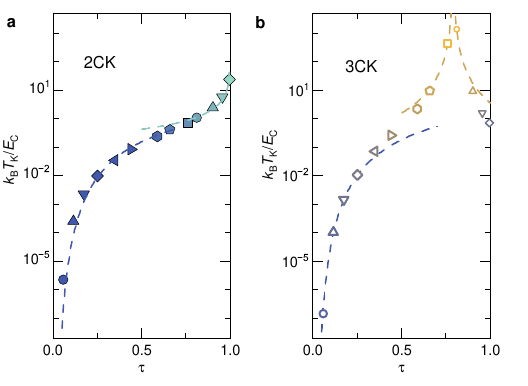}
\caption{
\footnotesize
\textbf{Kondo temperature evolution.} The experimentally-determined Kondo temperature $T_\mathrm{K}$ is displayed as symbols vs $\tau$, compared with theoretical predictions for tunnel contacts with $\tau \ll 1$ (blue dashed lines) and for cases of large $T_\mathrm{K}$ at $|\tau - \tau_c| \ll 1$ (green/orange dashed lines) for 2CK/3CK in a/b. 
}
\label{ExtFig4}
\end{figure}

\noindent\textbf{Kondo temperature.}
The Kondo temperature $T_\mathrm{K}$ is extracted for each QPC configuration by fitting the channel conductance at charge degeneracy (see Main Paper Figure~\ref{fig4}). In Extended Figure~\ref{ExtFig4} we plot the evolution  $T_\mathrm{K}(\tau)$ vs $\tau$, comparing with theory predictions. As previously observed\cite{Iftikhar2018}, $T_\mathrm{K}$ diverges near a critical transmission tuning $\tau_c$, with $\tau_c=1$ at 2CK (as specifically predicted, see Eq. (\eqref{eq:TKstar}) below) and $\tau_c\sim0.8$ at 3CK. Note that since $T_\mathrm{K}$ is a temperature scaling parameter, it is possible to have $k_\mathrm{B}T_\mathrm{K}$ much larger than the high energy cutoff $E_\mathrm{C}$ although, in such cases, only a small fraction $T/T_\mathrm{K}\ll E_\mathrm{C}/k_\mathrm{B}T_\mathrm{K}$ of the universal flow trajectory is accessible. 
Dashed lines display the predicted asymptotic behavior of the Kondo temperature in the weak coupling $\tau\ll 1$ limit and near the $\tau \to \tau_c$ limit\cite{Furusaki1995b,Iftikhar2018}:
\begin{eqnarray}
T_\mathrm{K}(E_\mathrm{C},\tau\ll1) &\propto& \frac{E_\mathrm{C}}{k_\mathrm{B}} \exp\left(-\frac{\pi^2}{2\sqrt{\tau}}\right),\\
T_\mathrm{K}(E_\mathrm{C},|\tau-\tau_c|\ll1) &\propto& \frac{E_\mathrm{C}}{k_\mathrm{B}} \frac{1}{|\tau - \tau_c|^a},
\end{eqnarray}
with the predicted exponent $a=1$ at 2CK and $a=5/2$ at 3CK.
The observed data-theory agreement further establishes the precise charge-Kondo implementation of the investigated device and the validity of the experimental procedures.\\

\noindent\textbf{Weak coupling predictions.}
When the metallic island is weakly coupled to the reservoirs via two tunnel contacts ($\tau\ll 1$), it effectively forms a single-electron transistor (SET). Such an effect applies whatever the number of channels involved, as long as they are all weakly coupled.
The mean excess island charge due to a detuning $\Delta E$, relative to the charge degeneracy point, is straightforwardly given by the Boltzmann population ratio:
\begin{equation}
\langle\delta N\rangle
=\frac{1}{2}\tanh\left(\frac{\Delta E}{2k_\mathrm{B}T}\right).
\label{eq:Charge_lowT}
\end{equation}
The predictions of Eq.~\eqref{eq:Charge_lowT} are displayed in Fig.~\ref{fig3}a. 
The entropy follows straightforwardly from Eqs.~\eqref{eq:Charge_lowT} and \eqref{eq:DeltaS}, yielding the predictions shown in Fig.~\ref{fig3}e.
Transport is also governed by thermally activated charge fluctuations between two adjacent island charge states.
In this sequential-tunnelling regime, the linear conductance across the island takes the form\cite{Nazarov_Blanter_2009}:
\begin{equation}
\frac{G_{\mathrm{SET}}(\delta V_{\mathrm{pl}},T)}{G_{\mathrm{SET}}(0,T)} =
\frac{2 E_\mathrm{C}}{k_\mathrm{B} T} \frac{\delta V_{\mathrm{pl}}}{\Delta} \Big/
\sinh\left[ \frac{2 E_\mathrm{C}}{k_\mathrm{B} T} \frac{\delta V_{\mathrm{pl}}}{\Delta} \right],
\label{eq:G_SET_lowT}
\end{equation}
where $G_{\mathrm{SET}}(0,T)$ is the conductance at charge degeneracy.
In Fig.~\ref{fig2}a, conductance data points at a relatively weak coupling, with a moderate $G_{\mathrm{SET}}(0,T)\sim0.3e^2/h$, cannot be directly confronted with the tunnel prediction given Eq.~\eqref{eq:G_SET_lowT} (the choice of rather weak and strong couplings displayed in Fig.~\ref{fig2}a was made to ensure a good visibility on a vertical axis extending over the full range). However, it is theoretically expected that first deviations to the peaks line shape Eq.~\eqref{eq:G_SET_lowT} by an increased coupling are captured by a reduction of the effective charging energy $E_\mathrm{C}^*<E_\mathrm{C}$\cite{Grabert1994}.
Accordingly the theoretical yellow lines in Fig.~\ref{fig2}a are obtained from Eq.~\eqref{eq:G_SET_lowT} fitting $G_{\mathrm{SET}}(0,T)$ and using a reduced $E_\mathrm{C}^* =28\,\mu$eV.

The universal evolution of the peak height due to Kondo renormalization has also been predicted in the limit of $T/T_\mathrm{K}\gg1$ ($\tau\ll1$) to scale as\cite{Furusaki1995b,Iftikhar2015}:
\begin{equation}
G_\mathrm{SET}(0,T) = G_{\mathrm{1,2(,3)}}^{\tau\ll1} \propto \frac{e^2}{h} \pi^2/\log^2(\alpha T/T_\mathrm{K})
\end{equation}
This asymptotic power law is represented as blue dashed lines in Fig.~\ref{fig2}b,c, with $\alpha$ a numerical factor depending on the precise definition of $T_\mathrm{K}$, and thus on the number of channels.\\

\noindent\textbf{Strong-coupling 2CK predictions.} 
In the near-ballistic regime ($1-\tau\ll1$) and at low temperatures ($k_\mathrm{B}T\ll E_\mathrm{C}$), the 2-channel charge Kondo effect can be quantitatively predicted based on Matveev’s model\cite{Matveev1995,Matveev1991}, as previously established in Refs.~\citenum{Iftikhar2015,Iftikhar2018,han_fractional_2022,piquard_observing_2023}. 
The predictions presented below are used to compare with the measured conductance, island charge and impurity entropy in Figs.~\ref{fig2}, \ref{fig3} and \ref{fig5}.
The conductance is given by\cite{Furusaki1995b,Iftikhar2015,Iftikhar2018}:
\begin{equation}
G_{\mathrm{1,2}}^{1-\tau\ll1}=\frac{e^2}{h}
\left[
1-T/T^*_\mathrm{K}-\displaystyle \int_0^{\infty} \frac{\cosh^\mathrm{-2} (x)  }{ 1 + (2x T / T_{\mathrm{co}})^2 }\mathrm{d}x
\right],
\label{eq:G2CK}
\end{equation}
where $T^*_\mathrm{K}$, identified with the Kondo temperature at charge degeneracy (up to a numerical factor that depends on convention) and the crossover temperature $T_{\mathrm{co}}$ are given by:
\begin{align}
T^*_\mathrm{K} &= \frac{4 E_\mathrm{C}/k_\mathrm{B}}{e^{C}\pi^3(1-\tau)\cos^2\left(\pi\frac{\delta V_{\mathrm{pl}}}{\Delta} \right)}, 
\label{eq:TKstar}\\
T_{\mathrm{co}} &= \frac{8 E_\mathrm{C} e^{C} (1 - \tau)}{k_\mathrm{B} \pi^2}
\sin^2 \left( \pi \frac{\delta V_{\mathrm{pl}}}{\Delta} \right),
\label{eq:Tco}
\end{align}
with $C \simeq 0.5772$ the Euler’s constant.  
When $T/T^*_\mathrm{K}$ is sufficiently small, $G_{\mathrm{2CK}}^{^{1-\tau\ll1}}$ solely depends on $T/T_{\mathrm{co}}$, in agreement with the generically expected universality of the crossover from quantum criticality near the quantum-critical point.

\begin{figure}[t]
\renewcommand{\figurename}{\textbf{Extended Figure}}
\renewcommand{\thefigure}{\textbf{\arabic{figure}}}
\centering\includegraphics [width=1\columnwidth]{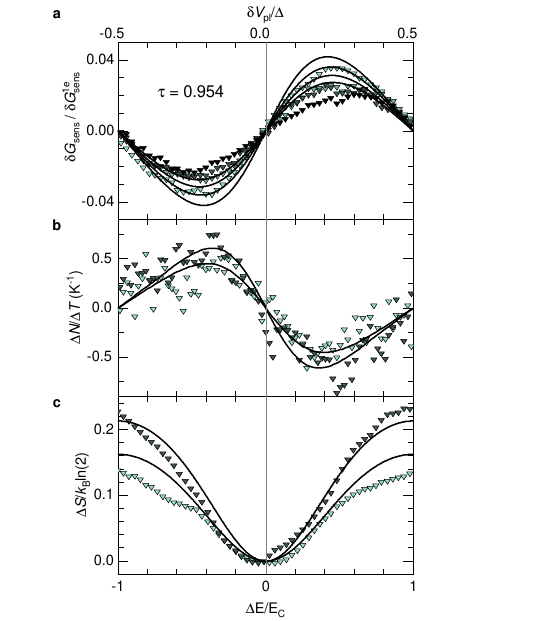}
\caption{
\footnotesize 
\textbf{Strong Coupling Predictions} are compared without adjustable parameters to the same $\tau = 0.954$ 2CK data shown in Fig.~\ref{fig3}b,d,f. 
        Sensor signal (a), temperature derivative (b), and extracted impurity entropy difference (c) are displayed versus $\Delta E/E_\mathrm{C} = \delta V_\mathrm{pl}/\Delta$. Symbols with different shades correspond to different temperatures, as in Fig.~\ref{fig3}. 
        Theoretical lines here are computed using the experimental value $\tau = 0.954$, in the absence of any fit parameter.
      }
\label{ExtFig5}
\end{figure}

The mean electron number in the island was first derived at $T=0$\cite{Matveev1995} and then extended to finite temperatures\cite{han_fractional_2022}:
\begin{equation}
\begin{aligned}
\langle N\rangle=
\frac{\delta V_{\mathrm{pl}}}{\Delta}
-\frac{2 e^{C} (1-\tau)}{\pi^2}
\Bigg[
1 &+ \ln \!\Bigg(
\frac{2\pi T}{T_{\mathrm{co}} }(1 - \tau) 
\sin^2 \!\left( \pi \frac{\delta V_{\mathrm{pl}}}{\Delta} \right)
\Bigg) \\
&+ \Psi^{(0)} \!\left( \frac{1}{2} + \frac{T_{\mathrm{co}}}{2\pi T} \right)
\Bigg]
\sin \!\left( 2\pi \frac{\delta V_{\mathrm{pl}}}{\Delta} \right),
\end{aligned}
\label{eq:FullCharge}
\end{equation}
where $\Psi^{(0)}$ is the di-gamma function.
From the above results, the impurity entropy difference $\Delta S$ along the crossover from quantum criticality at $T\ll T_\mathrm{K}$ (or equivalently $1-\tau\ll1$) is universal\cite{han_fractional_2022} and solely depends on the ratio $T/T_{\mathrm{co}}$:
\begin{equation}
\begin{aligned}
\Delta S 
=\frac{k_\mathrm{B}T_{\mathrm{co}}}{2\pi T} &\left[
\Psi^{(0)} \left( \frac{1}{2} + \frac{T_{\mathrm{co}}}{2 \pi T} \right) - 1
\right]\\
- k_\mathrm{B}\ln &\left[
\frac{1}{\sqrt{\pi}} \Gamma \left( \frac{1}{2} + \frac{T_{\mathrm{co}}}{2 \pi T} \right)
\right],
\end{aligned}
\label{eq:Entropy_2CK}
\end{equation}
where $\Gamma$ is the gamma function. The same crossover was obtained by completely different methods in the opposite (weak coupling) limit $\tau\ll1$ where the spin-$\tfrac{1}{2}$ impurity multi-channel Kondo model describes the physics, see e.g.~\citenum{mitchell2012universal}. 
All 2CK theoretical curves shown in Figs.~\ref{fig2}a, \ref{fig5}a,c as continuous black lines were computed from the above expressions without fit parameters, using experimentally determined $T$, $E_\mathrm{C}$, $\tau$, $\Delta$, and $\delta V_{\mathrm{pl}}$.\\
For the particular case of Fig.~\ref{fig3}b,d,f, the selected value of $\tau\simeq0.954$ is a compromise between sufficient signal to noise ratio on the charge and entropy data (requiring $1-\tau$ is not too small) and sufficiently strong coupling to accurately use Matveev predictions (small $1-\tau$).
In Fig.~\ref{fig3}b,d,f, we chose to use $\tau$ as a free parameter in the theory to account for the renormalization escaping the perturbative predictions, and found a best fit value of $\tau=0.966$.
For completeness, we also display in Ext.Fig.~\ref{ExtFig5} a direct, parameter-free comparison using $\tau=0.954$.\\

\noindent\textbf{Coulomb valley entropy in the strong-coupling 2CK regime.}
We detail here how are obtained the crosses shown in the 2CK entropy renormlization flow Fig.~\ref{fig4}a.
In the strong-coupling two-channel Kondo regime ($1-\tau\ll1$) where $T\ll T_\mathrm{K}$), the impurity entropy at maximum detuning ($\delta V_\mathrm{pl}=\Delta /2$) can be relatively important, which reduces the measured entropy difference $\Delta S(E_\mathrm{C})$.
In order to be able to compare with the predicted Kondo impurity entropy, the displayed crosses are obtained by adding to the measured $\Delta S$ the calculated Coulomb valley impurity entropy.
Specifically, $S_\mathrm{imp}^{\mathrm{thy}}(\delta V_\mathrm{pl}=\Delta/2)$ is straightforwardly obtained  from the universal 2CK crossover expressions Eqs.~\eqref{eq:Entropy_2CK} and \eqref{eq:Tco} valid at $1-\tau\ll1$.
The displayed Kondo impurity entropy at charge degeneracy is given by
\[
S_\mathrm{imp}(0)=\Delta S+S_\mathrm{imp}^{\mathrm{thy}}(\delta V_\mathrm{pl}=\Delta/2).
\]
This correction is applied only to the highest $\tau$, corresponding to $T/T_\mathrm{K} \lesssim 0.02,$ in Fig.~\ref{fig4}a, for which Matveev predictions are expected to be most accurate.
The resulting data points (cross symbols) yields values consistent with a low temperature saturation toward the universal 2CK entropy $k_\mathrm{B}\ln\sqrt{2}$. 
We emphasize that this procedure relies on a theoretical model that predicts Kondo criticality. 
Therefore, this comparison strengthen our conclusion and shows that the observed $\Delta S$ behavior follows theoretical predictions, but it should not be considered as a direct demonstration of a 2CK fractional entropy.
Note that no analogous compensation could apply in the three-channel Kondo case, for lack of theoretical predictions for the Coulomb valley impurity entropy in the strong-coupling 3CK regime.\\

\noindent\textbf{Frozen impurity criterion.}  
The thermodynamic Maxwell relation Eq.~\eqref{eq:DeltaS} informs on the difference between the Kondo impurity entropy of interest ($S_\mathrm{imp}(0)$, at charge degeneracy $\delta V_\mathrm{pl}=0$) and the residual impurity entropy in the Coulomb valley ($S_\mathrm{imp}(\Delta E)$, with $\Delta E=2E_\mathrm{C}\delta V_\mathrm{pl}/\Delta$).
This relation directly provides $S_\mathrm{imp}(0)$ only when it is possible to separately ascertain that $S_\mathrm{imp}(\Delta E)\approx0$.
In practice, we use the conductance across the device (through the island), and more precisely its reduction $R$ in the middle of the Coulomb valley ($\delta V_\mathrm{pl}=\Delta/2$) with respect to its value at charge degeneracy (see insets in Fig.~\ref{fig4}):
\[
R \equiv \frac{G_i(\delta V_\mathrm{pl}=\Delta/2)}{G_i(0)}.
\]
Qualitatively, a small $R$ indicates a freezing of the island's charge at $\delta V_\mathrm{pl}=\Delta/2$, and thus a small $S_\mathrm{imp}(\Delta E=E_\mathrm{C})$.
Quantitatively, a direct relation can be most straightforwardly established in the weak-coupling limit ($\tau\ll1$), from Eqs.~\eqref{eq:Charge_lowT} and \eqref{eq:G_SET_lowT}.
Such a strong relation between $R$ and charge pinning (and entropy) extends to strong couplings ($1-\tau\ll1$), as can be shown from Eqs.~\eqref{eq:G2CK}, \eqref{eq:FullCharge} and \eqref{eq:Entropy_2CK}. 
The connection between $R$ and charge freezing at $\delta V_\mathrm{pl}$ forms the backbone of a previous experimental study of charge quantization\cite{Jezouin2016}.
In practice, we adopt the criterion $R<0.075$, for which both weak‑coupling and strong-coupling predictions consistently attest that $S_\mathrm{imp}(\Delta E=E_\mathrm{C})\lesssim0.1\,k_\mathrm{B}\ln2$.
Obeying this criterion therefore ensures that the residual contribution of $S_\mathrm{imp}$ in the Coulomb valley to the measured $\Delta S(\Delta E=E_\mathrm{C})$ is smaller than the experimental accuracy and can be ignored.
The $\Delta S$ data verifying this criterion are shown as colored symbols in Fig.~\ref{fig4}.
These could be directly compared to the NRG prediction for $S_\mathrm{imp}(0)$ (solid line).
The data shown as grey symbols in Fig.~\ref{fig4} do not fulfill $R<0.075$. Therefore, the measured $\Delta S$ is expected (and observed) to be significantly reduced compared to the Kondo impurity entropy, with stronger reductions for higher $R$ (see inset).\\

\begin{figure}[t]
\renewcommand{\figurename}{\textbf{Extended Figure}}
\renewcommand{\thefigure}{\textbf{\arabic{figure}}}
\centering\includegraphics [width= 85mm]{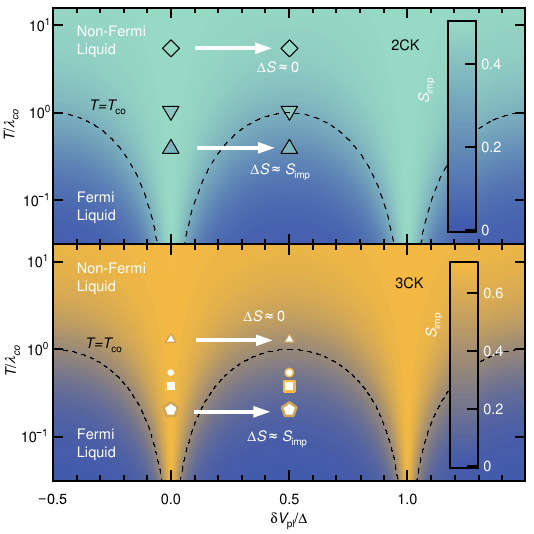}
\caption{
\footnotesize
\textbf{Crossover from quantum criticality.} 
Schematic colormap of the expected impurity entropy $S_\mathrm{imp}$ as a function of both $\delta V_\mathrm{pl}/\Delta$ and $\lambda_\mathrm{co}/T$ for 2CK (a) and 3CK (b). Symbols show the experimental ratios of $\lambda_\mathrm{co}/T$ at the lowest temperature used for entropy measurement, for various transmissions $\tau$. The measured entropy $\Delta S$ corresponds to the difference between the impurity entropy at the degeneracy point ($\delta V_\mathrm{pl}=0$) and in the Coulomb valley ($\delta V_\mathrm{pl}=\Delta/2$). Dashed lines represent the frontier $T_\mathrm{co}=T$.}
\label{ExtFig6}
\end{figure}

\noindent\textbf{Entropy along crossover from quantum criticality.}
Here we present the general picture of the crossover away from 2CK or 3CK quantum criticality induced by detuning the device away from charge degeneracy.
Starting from a well-developed quantum criticality ($T\ll T_\mathrm{K}$ here), any small detuning away from the quantum-critical point (from charge degeneracy here) gives rise to a crossover away from quantum criticality as temperature is further reduced.
In Ext.\ Fig.~\ref{ExtFig6}, this would correspond to following a vertical line (detuned from $\delta V_\mathrm{pl/\Delta}\in\mathbb{Z}$) starting from a non-Fermi liquid state at higher $T$ to enter a Fermi liquid state for lower $T$.
This crossover develops around a characteristic temperature scale $T_\mathrm{co}$ which depends on the detuning.
Importantly, $T_\mathrm{co}$ is generically expected to encapsulate all microscopic details, such that the crossover from quantum criticality is a universal function of $T/T_\mathrm{co}$.
Here, $T_\mathrm{co}$ is given by Eq.~\eqref{eq:Tco_main}.
It is a periodic function of the detuning $\delta V_\mathrm{pl}$, as shown by the corresponding dashed lines in Ext.\ Fig.~\ref{ExtFig6}.
The universal $T/T_\mathrm{co}$ character of the crossover from quantum criticality allows us to explore it not only by changing the temperature $T$ at a fixed detuning (fixed $T_\mathrm{co}$), but also by changing the detuning (and hence $T_\mathrm{co}$) at a fixed $T$.
The second approach is the one used in the main manuscript to establish a lower bound for the Kondo impurity entropy (see Fig.~\ref{fig5}).
The corresponding crossover path is illustrated by horizontal white arrows in Ext.\ Fig.~\ref{ExtFig6}.
Because the $\delta V_\mathrm{pl}$ periodicity effectively limits the increase of $T_\mathrm{co}$, the crossover can only be explored at fixed $T$ down to the corresponding minimum value of $T/T_\mathrm{co}$ (reached at $\delta V_\mathrm{pl}=\Delta/2$).
For the lower arrow in the bottom (3CK) panel, this minimal value is essentially within the Fermi-liquid state and, correspondingly, a large fraction of the crossover can be explored.
In contrast, the higher white arrows correspond to maximum values of $T_\mathrm{co}$ that remain higher than $T$ and, consequently, only a relatively small fraction of the crossover is accessible by detuning $\delta V_\mathrm{pl}$.
The crossover paths explored in Fig.~\ref{fig5} are indicated by the corresponding symbols in Ext.\ Fig.~\ref{ExtFig6}, on top of a colormap corresponding to the entropy computed analytically for 2CK (top panel, Eq.~\eqref{eq:Entropy_2CK}) and by NRG for 3CK (bottom panel).\\

\noindent\textbf{Universal crossover criterion.}
We address here the criterion used to select the data shown in Fig.~\ref{fig5} for exploring the universal $T/T_\mathrm{co}$ crossover from quantum criticality.
Universality requires that the starting, quantum critical point is sufficiently deep into quantum criticality.
In the present 2CK and 3CK cases, this corresponds to data points of low $T/T_\mathrm{K}$ (at charge degeneracy where Kondo physics fully develops).
Pragmatically, we focus on the observable of interest (entropy), and require that, at the charge degenerate critical point, the Kondo entropy deviates from the 2CK or 3CK quantum critical value by less than our $0.1\,k_\mathrm{B}\ln2$ experimental uncertainty.
From NRG prediction, this is achieved (for both 2CK and 3CK) provided the minimal criterion $T/T_\mathrm{K}<0.02$ is verified.
Figure~\ref{fig5} shows all the explored device settings found to match this criterion.\\

{\noindent\textbf{NRG calculations.}}
The numerical renormalization group method\cite{Wilson1975}, exploiting the interleaved Wilson chain representation\cite{Mitchell2014}, was used to compute the universal curves for conductance and impurity entropy in both 2CK and 3CK settings of the multi-channel Kondo model, Eq.~\eqref{eq:MCKmodel}.  Conductance was computed in the linear-response regime using a generalized Kubo formula\cite{minarelli2022linear}, as also employed in Refs.~\citenum{Iftikhar2018,han_fractional_2022}. The large transmission regime of the charge-Kondo system, where many island charge states contribute, lies beyond the reach of NRG techniques at present. However, the universality of the 2CK and 3CK Kondo critical points means that the entire Fermi-liquid crossover\cite{mitchell2012universal} is exactly described by the NRG solution of Eq.~\eqref{eq:MCKmodel} for any coupling strength $\tau$, provided $T\ll \min(T_\mathrm{K}, E_\mathrm{C})$. \\

\begin{figure}[t]
\renewcommand{\figurename}{\textbf{Extended Figure}}
\renewcommand{\thefigure}{\textbf{\arabic{figure}}}
\centering\includegraphics [width= 85mm]{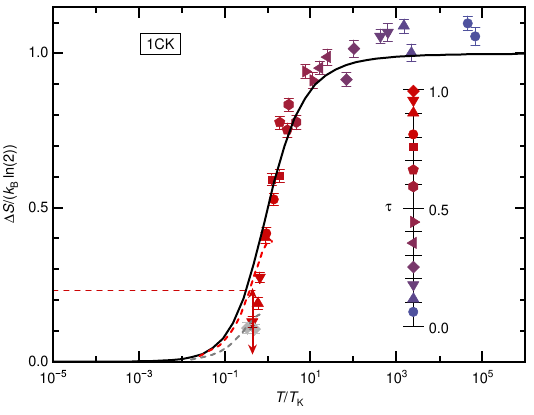}
\caption{
\footnotesize
\textbf{Entropy renormalization flow at 1CK.} 
    The universal NRG predictions for $S_\mathrm{imp}(0)$ is shown as a solid line.
    Colored symbols represent experimental data points for which the measured $\Delta S$ is predicted to closely approximate the Kondo impurity entropy $S_\mathrm{imp}(0)$.
    The horizontal dashed line in the main panel represents the associated upper bound for the Kondo ground state entropy, taking into account the estimated $\Delta S$ uncertainty of $\pm0.1\,k_\mathrm{B}\ln2$ ($\updownarrow$). The smaller instrumental standard error is displayed as error bars when bigger than symbols.
    Grey symbols correspond to the highest transmission tuning $\tau\simeq 0.993$ for which the Coulomb valley entropy is predicted to significantly reduce $\Delta S$. 
    The corresponding, non-universal analytical prediction for the measured $\Delta S$ is shown as a grey dashed line. 
    In contrast, the same theory for the slightly lower transmission setting $\tau\simeq0.954$ predicts that $\Delta S$ at 1CK is already very close to the universal Kondo entropy (colored dashed line).
    Symbols shape and size match those for identical configurations in Fig.~\ref{fig2}.
}
\label{ExtFig7}
\end{figure}

{\noindent\textbf{Entropy renormalization flow at 1CK.}}
Extended Fig.~\ref{ExtFig7} shows as symbols the measured entropy difference  $\Delta S$ between charge degeneracy and Coulomb valley, as a function of $T/T_\mathrm{K}$, in the presence of a single Kondo channel. It complements the figure~\ref{fig4} focused on 2CK and 3CK.
In practice, this regime is achieved by connecting only one QPC to the metallic island, and by setting it to transmit a single channel (the right QPC in Fig.~\ref{fig1}) with a probability $\tau$, while the others are disconnected ($\tau\equiv\tau_1$, $\tau_2=\tau_3=0$).\\
As there is only one connected channel, the conductance across the QPC through the island cannot be measured in-situ, which implies a different procedure to set $T_\mathrm{K}$ and to signal that the charge pseudo spin in the Coulomb valley is not frozen. 
For the former, we use here the charge (Kondo pseudospin) susceptibility at charge degeneracy. Following our previous work Ref.~\citenum{piquard_observing_2023}, we set $T_\mathrm{K}$ from the comparison between data and predicted renormalization flow for the pseudospin susceptibility at 1CK. 
Then this value of $T_\mathrm{K}$ is used to make a parameter-free comparison between entropy data and NRG predictions (continuous black line).
For the valley frozen criteria, we resorted to the entropy difference predictions obtained within Matveev formalism for $1-\tau\ll1$ (applying Eq.~\ref{eq:DeltaS} on the 1CK charge predictions at $1-\tau\ll1$ from Ref.~\citenum{piquard_observing_2023}, Eq. 4), and compared these predictions with the NRG computations for $S_\mathrm{imp}$.
Colored symbols indicate an essentially frozen charge in the Coulomb valley, corresponding in practice to a Matveev-NRG difference below $0.075  k_\mathrm{B}\ln2$.
Grayed symbols indicate device settings (in practice only the highest  transmission $\tau\simeq 0.993$) for which we expect that the charge is not frozen in the valley, and therefore that the measured entropy difference is significantly reduced compared to the Kondo entropy at charge degeneracy.
In the present 1CK case there is no universal crossover from quantum criticality, therefore there is no equivalent of Eqs.~\ref{eq:Tco},~\ref{eq:Entropy_2CK} used at 2CK to obtain the Coulomb valley entropy.
Instead of using a less straightforward estimate, we chose to display the quantitative, non-universal predictions for the measured $\Delta S$ which is given analytically within Matveev framework at $1-\tau\ll1$. 
These (non-universal) quantitative predictions, obtained without any adjustable parameters, are shown as dashed lines of the same color as the corresponding symbols.\\
This data-theory comparison establishes the predicted entropy renormalization flow at 1CK down to $T/T_\mathrm{K}\simeq0.04$. Deviations at lower $T/T_\mathrm{K}$ are accounted for quantitatively by the predicted $\Delta S$.
Including the estimated experimental uncertainty of $0.1 k_\mathrm{B}\ln2$ on $\Delta S$, the same procedure used for 2CK and 3CK would provide an experimental upper bound of $\simeq0.23 k_\mathrm{B}\ln2$ for the 1CK entropy.\\

{\noindent\textbf{Data availability.}}
The experimental data used to produce the plots in this paper are available as Source Data Files, provided with this paper.

\vfill

\bibliography{biblio}
\bibliographystyle{nature}

\vspace{\baselineskip}
\small
{\noindent\textbf{Acknowledgments.}}
This work was supported by the European Research Council (ERC-2020-SyG-951451). 
Nanofabrication was done within the C2N micro nanotechnologies platform and partly supported by the French RENATECH network and the general Council of Essonne. AKM acknowledges funding from Research Ireland through grant 21/RP2TF/10019. 
We thank K.~Ensslin, J.~Folk, T.~Ihn, C.~Mora, Y.~Meir and E.~Sela for discussions.\\

{\noindent\textbf{Author Contributions.}}
C.P.\ performed the experiment with input from A.Aa., A.An., A.V., F.P., F.Z.\ and Y.S.;
C.P., A.An.\ and F.P.\ analyzed the data;
C.P.\ and Y.S.\ fabricated the sample;
A.C.\ and U.G.\ grew the 2DEG;
A.K.M.\ provided theory support and did the NRG calculations;
C.P., A.K.M.\ and F.P.\ wrote the paper with input from all authors;
A.An.\ and F.P.\ led the project.\\

{\noindent\textbf{Competing interests}}
The authors declare no competing financial interests.\\

{\noindent\textbf{Author Informations}}
Correspondence and requests for materials should be addressed to C.P.\ (colin.piquard@espci.org), A.An.\ (anne.anthore@c2n.upsaclay.fr) and F.P.\ (frederic.pierre@cnrs.fr).\\
\normalsize

\end{document}